\documentclass[12pt, draftclsnofoot, onecolumn]{IEEEtran}
\IEEEoverridecommandlockouts
\usepackage{amsmath,amsfonts}
\usepackage{array}
\usepackage[caption=false,labelfont=sf,textfont=sf]{subfig}
\usepackage{textcomp}
\usepackage{stfloats}
\usepackage{url}
\usepackage{verbatim}
\usepackage{graphicx}
\usepackage{cite}

\usepackage{schemabloc,color,xspace}
\usepackage[ruled,vlined,linesnumbered,noend]{algorithm2e}
\usepackage{enumitem}

\usepackage{tabularx}
\usepackage{booktabs}
\definecolor{yellowp}{rgb}{0.80,0.7504,0.2294}
\definecolor{greenp}{rgb}{0.0,0.750,0.0}
\definecolor{redp}{rgb}{0.750,0.0,0.0}
\definecolor{bluep}{rgb}{0.0,0.0,0.75}
\definecolor{cyanp}{rgb}{0,1,1}
\definecolor{magentp}{rgb}{1,0.0,1}

\newcommand{\unsaturated}{U}
\newcommand{\disconnected}{D}
\newcommand{\scratch}{s}

\newcommand{\route}{r}
\newcommand{\error}{\delta}

\newcommand{\modulation}{M}
\usepackage{booktabs}
\newtheorem{theorem}{Theorem}
\newtheorem{proposition}[theorem]{Proposition}
\newtheorem{remark}[theorem]{Remark}
\newtheorem{definition}[theorem]{Definition}

\begin{document}
%
\title{Switching in the Rain: Predictive Wireless x-haul Network Reconfiguration}
\author{%
Igor Kadota\IEEEauthorrefmark{1}\IEEEauthorrefmark{2}, Dror Jacoby\IEEEauthorrefmark{1}\IEEEauthorrefmark{3}, Hagit Messer\IEEEauthorrefmark{4}, Gil Zussman\IEEEauthorrefmark{2}, and Jonatan Ostrometzky\IEEEauthorrefmark{3}
\thanks{\IEEEauthorrefmark{1}Both authors contributed equally to this research.}
\thanks{\IEEEauthorrefmark{2}I. Kadota and G. Zussman are with the Department of Electrical Engineering, Columbia University, New York, USA.}
\thanks{\IEEEauthorrefmark{3}D. Jacoby and J. Ostrometzky are with the Faculty of Engineering, Tel~Aviv University, Tel Aviv, Israel.}
\thanks{\IEEEauthorrefmark{4}H. Messer is with the School of Electrical Engineering, Tel~Aviv University, Tel Aviv, Israel.}
}

\maketitle



\begin{abstract}
Wireless x-haul networks rely on microwave and millimeter-wave links between 4G and/or 5G base-stations to support ultra-high data rate and ultra-low latency. A major challenge associated with these high frequency links is their susceptibility to weather conditions. In particular, precipitation may cause severe signal attenuation, which significantly degrades the network performance. In this paper, we develop a Predictive Network Reconfiguration (PNR) framework that uses historical data to predict the future condition of each link and then prepares the network ahead of time for imminent disturbances. The PNR framework has two components: (i) an Attenuation Prediction (AP)  mechanism; and (ii) a Multi-Step Network Reconfiguration (MSNR) algorithm. The AP mechanism employs an encoder-decoder Long Short-Term Memory (LSTM) model to predict the sequence of future attenuation levels of each link. The MSNR algorithm leverages these predictions to dynamically optimize routing and admission control decisions aiming to maximize network utilization, while preserving max-min fairness among the base-stations sharing the network and preventing transient congestion that may be caused by re-routing. We train, validate, and evaluate the PNR framework using a dataset containing over 2 million measurements collected from a real-world city-scale backhaul network. The results show that the framework: (i) predicts attenuation with high accuracy, with an RMSE of less than $0.4$ dB for a prediction horizon of $50$ seconds; and (ii) can improve the instantaneous network utilization by more than $200\%$ when compared to reactive network reconfiguration algorithms that cannot leverage information about future disturbances 

\end{abstract}







%



\section{Introduction}

\begin{figure}[t]
\centering
\subfloat[City-scale backhaul network]
{\includegraphics[width=0.3\columnwidth]{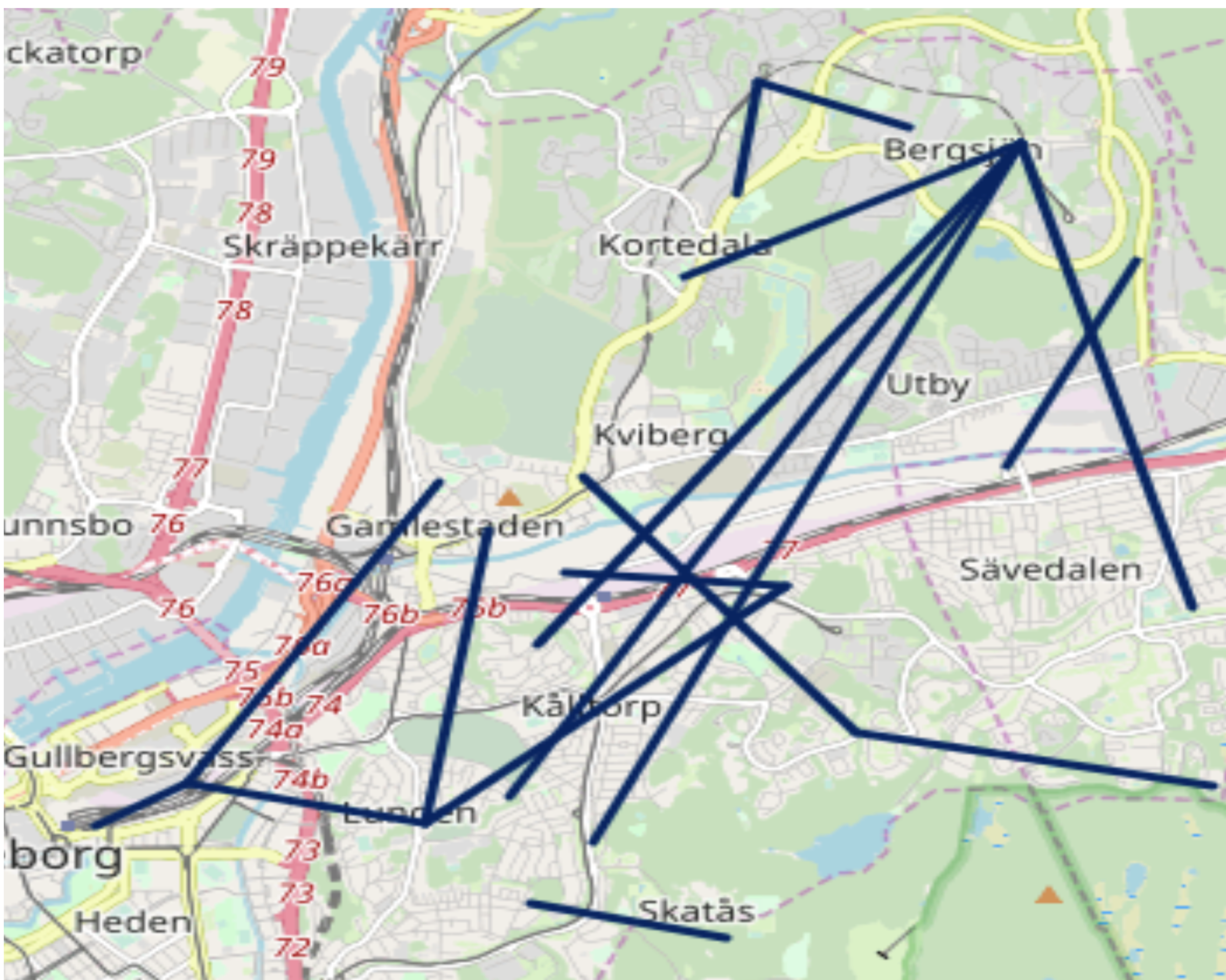}}
\hspace{0.5cm}
\subfloat[Network Abstraction]
{\includegraphics[width=0.3\columnwidth]{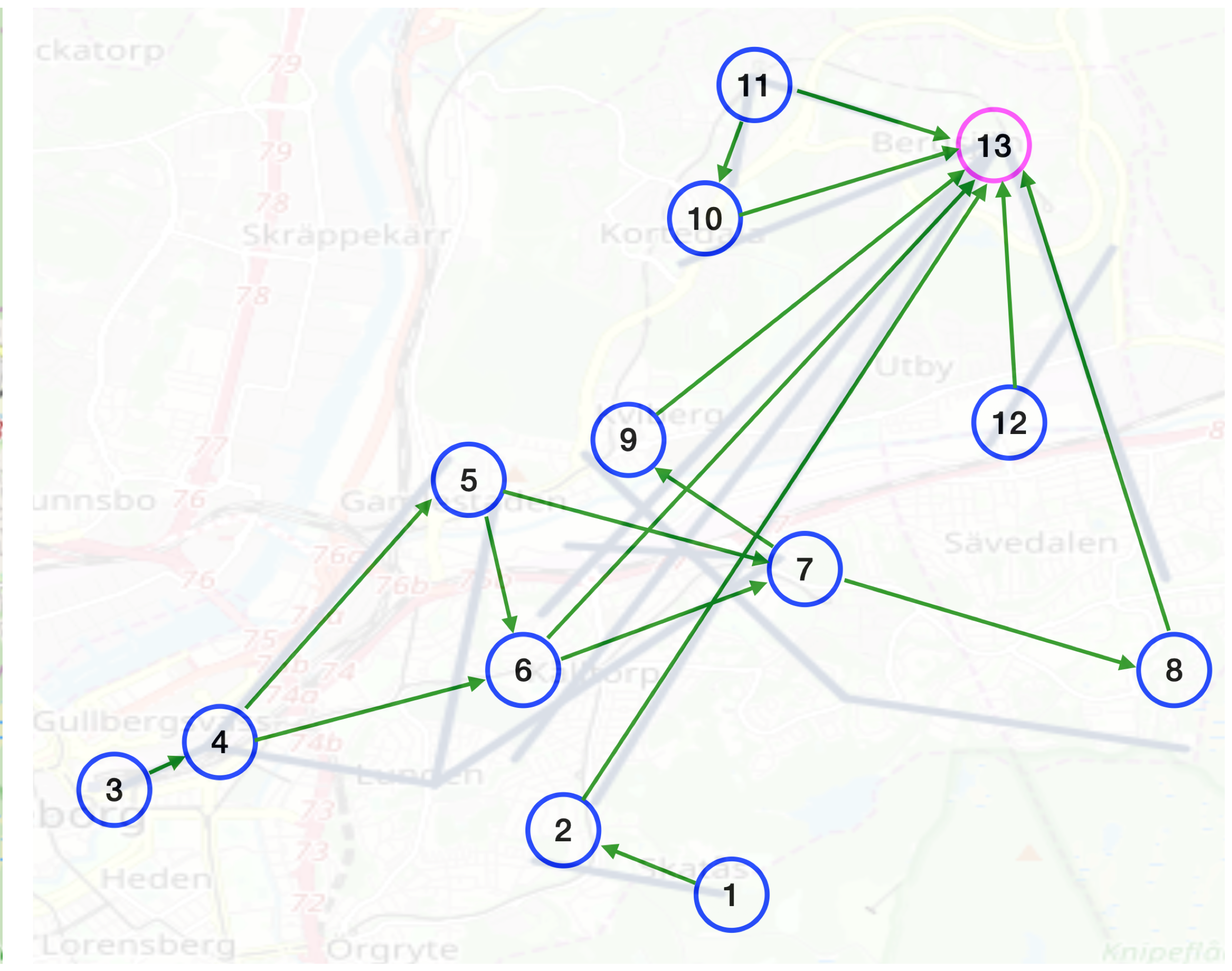}}
\vspace{-0.2cm}
\caption{(a) A wireless backhaul network in Gothenborg, Sweden (the map area is of approximately 10x10 km$^2$). The data utilized in this paper was collected from this network by \emph{Ericsson AB}. (b) An abstraction of the network topology (described in Sec.~\ref{sec.Model}).} \label{CMLSweden}
\vspace{-0.4cm}
\end{figure}

\begin{figure}[t]
\centering
\includegraphics[width=0.4\columnwidth]{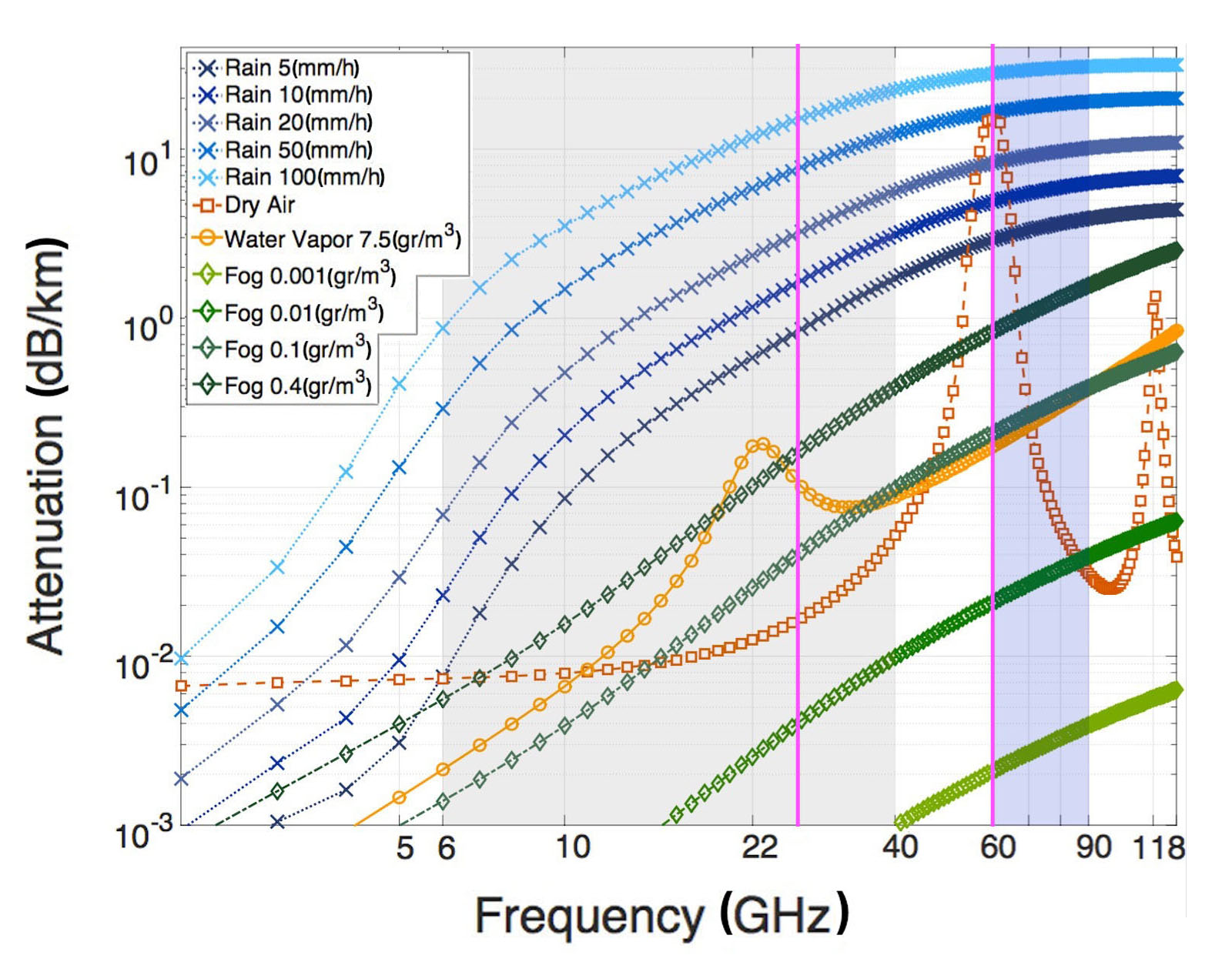}
\vspace{-0.2cm}
\caption{Signal attenuation (in dB/km) for various atmospheric phenomena as a function of frequency~\cite{ITU530}. The commonly used bands of 6--40 GHz (K-band) and 60--90 GHz (E-band) are highlighted.} \label{fITU}
\vspace{-0.5cm}
\end{figure}

4G and 5G networks often use high bandwidth microwave and millimeter-wave (mmWave) links in their fronthaul, midhaul, and backhaul (x-haul) networks~\cite{Jonas2017} for supporting applications that require high data rate and ultra-low latency. These wireless x-haul networks can connect a large number of base-stations, covering entire cities, as depicted in Fig.~\ref{CMLSweden}(a). A main challenge of using microwave and mmWave links is their high susceptibility to weather conditions. The signal attenuation due to different atmospheric and weather phenomena is described by the International Telecommunication Union (ITU) in \cite{ITU530,ITU840,ITU,ITU676} and depicted in Fig.~\ref{fITU}. It can be seen that, apart from the oxygen resonance frequency at $60$~GHz, the dominant factor affecting link attenuation is precipitation. This implies that signal attenuation may vary significantly over time and over geographic locations. \emph{Hence, the need for a high bandwidth wireless x-haul that is robust to variations in the network conditions calls for the development of a predictive network reconfiguration framework that can dynamically allocate resources based on current and future estimated network conditions}. 


Until recently, only local Physical/Link layer mechanisms were employed to alleviate the impact of the time-varying conditions of the links on the network performance. For example, the Automatic Transmit Power Control is a commonly used mechanism that adjusts the transmitter power based on measurements of the link attenuation \cite{Goldsmith2001WirelessModulation}. However, with the emergence of Software-Defined Networking (SDN)~\cite{Caesar2005,McKeown2008,Jain2013}, it is now possible to develop global Network layer mechanisms (such as NEC's backhaul solution in \cite{Necsol}) that monitor the entire network and react to performance drops caused by weather-induced disturbances. A main drawback of \emph{reactive reconfiguration} mechanisms is their delay in recovering from performance drops, which may severely affect time-sensitive applications. To overcome this challenge, \emph{predictive reconfiguration} mechanisms can be employed. 

Prior work on predictive network reconfiguration algorithms (see \cite{bui2017survey} for a survey) focused mainly on alleviating the effects of node mobility \cite{Naimi2014,Lu2013,margolies_exploiting_2016,chattopadhyay2018location,yang2015sensor,doff2015sensor}, traffic demand variability \cite{Huang2014,Abedini2014,Balachandran2013,Papagiannaki2003,Sun2016,wang2018spatio,bega2019deepcog,Wang2013Crowded}, and link quality degradation due to multi-path reflection, line-of-sight occlusion, and interference \cite{Xiaozheng2011,Yin2011,Bui2015,Mangla2016,Tarsa2015,yue2018linkforecast,sur2016beamspy,Wang2015MAGNUS}. Weather effects pose fundamentally different challenges. In particular, weather-induced attenuation can be severe, affect large contiguous geographic areas, and last for extended periods of time. The literature on the prediction of microwave and mmWave signal attenuation due to weather conditions uses meteorological data (e.g., weather-radar echo measurements) to predict the current/future attenuation levels~\cite{jabbar2009performance,Rak2016} or uses past attenuation measurements to predict future attenuation levels~\cite{Yaghoubi2018,jacoby2020short,jacoby2021, patel2014implementation, 875201}. 
%
Most relevant to this paper is our prior work in~\cite{jacoby2020short} which employs an encoder-decoder LSTM model to predict future link attenuation levels. The main drawbacks of~\cite{jacoby2020short} are that: (i) its prediction mechanism does not capture the significant spatial correlation of the rain-induced attenuation; and (ii) its prediction mechanism is not employed to inform any algorithm or protocol.

The literature on predictive weather-aware reconfiguration algorithms contains only a few works \cite{jabbar2009performance,Rak2016,Javed2013,Yaghoubi2018}. Most of these works, in particular \cite{jabbar2009performance,Rak2016,Javed2013}, develop modifications to standard \emph{distributed} routing protocols such as Open Shortest Path First (OSPF) which due to the lack of centralized coordination may converge slowly, making them unsuitable for networks that support time-sensitive applications. Only \cite{Yaghoubi2018} leverages SDN to perform centralized predictive network-wide reconfiguration. The framework proposed in \cite{Yaghoubi2018} predicts future link attenuation levels using a model \emph{specific to rain fading} and then computes current and future routing decisions aiming to maximize throughput. Some limitations of the solution proposed in \cite{Yaghoubi2018} are that: (i) its attenuation prediction mechanism does not capture the spatial correlation of the weather-effects; (ii) its prediction mechanism can \emph{only} be employed during periods of rain; (iii) its network reconfiguration mechanism \emph{allows transient link congestion} (i.e., it allows flows to temporarily exceed the link capacity); (iv) its network reconfiguration mechanism does not take fairness into account; and (v) its re-routing mechanism does not support flow splitting. 

\textbf{Our contributions}: In this paper, we develop and evaluate, based on a real dataset, a Predictive Network Reconfiguration (PNR) framework that leverages existing local Physical/Link layer mechanisms and adds two new components: an Attenuation Prediction (AP) mechanism; and a Multi-Step Network Reconfiguration (MSNR) algorithm. 

\emph{The AP mechanism} employs an encoder-decoder LSTM model to predict the sequence of future attenuation levels based on past measurements, capturing both time and spatial correlation that are typical of weather-effects \emph{without incorporating weather-related models}, which allows it to be used both in dry and rain periods, and \emph{without relying on meteorological data from external sources} such as weather radars. To train, validate, and evaluate the AP mechanism, we use \emph{a unique dataset obtained from the real-world city-scale backhaul network in Gothenborg, Sweden} (see Fig.~\ref{CMLSweden}(a)) \emph{collected by Ericsson AB}. The dataset contains 2,295,000 measurements of link attenuation. In Fig.~\ref{fig.APmechanism}(a), we display the evolution of the measured attenuation for every link in the backhaul network over a period of $1.9$ hours. Notice that in the interval between $t=300$ and $600$ time-steps there is an increased attenuation due to rain. The spatio-temporal correlation is evident. \emph{The AP mechanism leverages this correlation to achieve high prediction accuracy}. In particular, the AP mechanism achieves a 
Root Mean Square Error (RMSE) of less than $0.4$\thinspace{dB} for a prediction horizon of $50$\thinspace{seconds}. We evaluate two benchmark time series prediction methods that do not capture the spatial correlation of the weather-effects and show that both of them can perform $30\%$ worse than the AP mechanism in terms of RMSE.



\emph{The MSNR algorithm} leverages the predictions from the AP mechanism and uses Model Predictive Control (MPC) \cite{MPC} to compute the sequence of current and future routing and admission control decisions that: (i) maximize network utilization, while (ii) achieving max-min fairness among the base-stations sharing the network and (iii) preventing transient congestion that may be caused by re-routing. This sequence of routing  and admission control decisions are employed by the centralized SDN controller to reconfigure the network over time. For example, based on a prediction that a set of links will become unavailable in $30$ seconds, the MSNR algorithm can determine when it is optimal for the SDN controller to redirect flows in order to avoid potential interruptions to service and can decide whether or not it is necessary to revoke network slices from low priority services. An important challenge associated with the MSNR algorithm is computational complexity. In Sec.~\ref{sec.Multple}, we proposed a principled implementation of the MSNR algorithm which has a computational complexity that grows polynomially with the prediction horizon, as opposed to a naive implementation that can have exponential complexity. 

 

\emph{We evaluate the PNR framework} using the data collected from the backhaul network. Our results show that the PNR framework can improve the instantaneous network utilization by more than $200\%$ when compared to reactive network reconfiguration algorithms that do not prepare the network for future disturbances. 
\emph{To the best of our knowledge, this is the first attempt to propose and evaluate, based on a real dataset, an integrated framework for x-haul network reconfiguration that leverages the spatio-temporal correlation of the weather-effects to jointly optimize routing and admission control decisions.} A patent including some of the results is pending \cite{patent}.

This paper is organized as follows.
Section~\ref{sec.Model} describes the network model and the dataset.
In Sec.~\ref{sec.APmechanism}, we develop the AP mechanism. 
In Sec.~\ref{sec.NRalgorithm}, we develop the MSNR algorithm. 
In Sec.~\ref{sec.Evaluation}, we evaluate the performance of the PNR framework. 
Section~\ref{sec.Conclusion} concludes the paper and discussed future work.

\section{Problem Formulation and Dataset}\label{sec.Model}

In this section, we present the network model used to develop the PNR framework. We first describe the model in general and then establish the connection between the model and the real-world backhaul network. Let $G=(V,E)$ be the directed graph that represents an x-haul communication network with base-stations, also called nodes, $n \in V=\{1$, $2,$ $\ldots,N\}$, connected by wireless links $(k,l)\in E$ where $k,l\in V$ and $(k,l)$ represents the link $k\rightarrow l$. Time is divided into time-steps with index $t\in\{1,2,\ldots,T\}$, where $T$ is the time-horizon and the time interval between $t$ and $t+1$ is $\Delta=10$ seconds. Let $d_n>0$ be the demand associated with commodity $n \in V$. The demand $d_n$ represents the uplink traffic that base-station $n$ aggregates from its associated users. Let $z_{n,t}\in[0,1]$ be the fraction of the demand $d_n$ admitted during time-step~$t$. It follows that the \emph{admitted demand} from base-station $n$ during time~$t$ is given by $z_{n,t}d_n$. For simplicity, we assume that demands $d_n$ remain fixed over time and that node $N$ is the common destination for all commodities $n \in V \setminus N$. Naturally, for the common destination $N$, we have $d_N=0$ and $z_{N,t}=0,\forall t$. Let $f_{n,t}^{(k,l)}\in[0,1]$ be the fraction of the admitted demand $z_{n,t}d_n$ that flows through link $(k,l)$ during time~$t$. By definition
\begin{align}
&f_{n,t}^{(k,n)}=0, \forall (k,n)\in E, \forall t\; ; \label{eq.flow_1}\\
&f_{n,t}^{(N,l)}=0, \forall n \in V, \forall (N,l)\in E, \forall t\; ; \label{eq.flow_2}\\
&f_{n,t}^{(k,l)}=0, \forall n \in V, \forall (k,l)\notin E, \forall t\; , \label{eq.flow_3}
\end{align}
where \eqref{eq.flow_1} is a constraint on the incoming flows at the source nodes, \eqref{eq.flow_2} is a constraint on the outgoing flows at the destination node $N$, and \eqref{eq.flow_3} enforces zero flow on non-existing links. 
It follows that the \emph{total flow} in link $(k,l)\in E$ during time~$t$ is given by $\sum_{n=1}^{N-1} z_{n,t}d_n f_{n,t}^{(k,l)}$. 

\textbf{Feasibility and Fairness.} We assume that $G=(V,E)$ and $d_n$ are given and remain fixed over time. We assume that routing and admission control decisions implemented by the centralized SDN controller at time-step~$t$, namely  $f_{n,t}^{(k,l)}$ and $z_{n,t}$, respectively, remain fixed in the interval between $t$ and $t+1$. 
Routing and admission control decisions at each time-step~$t$ are \emph{feasible} when they satisfy flow conservation and capacity constraints. 
The \emph{flow conservation} associated with commodity $n\in V$ and node $l\in V$ at time~$t$ is given by 
\begin{equation} \label{eq.flow}
\sum_{k=1}^{N}{ f_{n,t}^{(k,l)}}-\sum_{m=1}^{N}{ f_{n,t}^{(l,m)}} = 
    \begin{cases}
    -1,{\text{~}} l=n \\
    +1,{\text{~}} l=N \\
    0, {\text{~otherwise}}  
    \end{cases},
\end{equation}
where $l=n$ indicates that node~$l$ is the source of commodity~$n$ and $l=N$ indicates that node~$l$ is the destination of commodity~$n$. Let $c_t^{(k,l)}\geq 0$ be the capacity of link $(k,l)$ at time~$t$ and let $\hat{c}_{t+1}^{(k,l)}\geq 0$ be the \emph{predicted} capacity of link $(k,l)$ at time~$t+1$. Since the exact moment between $t$ and $t+1$ in which the capacity changes from $c_t^{(k,l)}$ to $\hat{c}_{t+1}^{(k,l)}$ is unknown, we assume the worst-case and represent the capacity in this interval 
by $\min\{c_t^{(k,l)},\hat{c}_{t+1}^{(k,l)}\}$. 
Hence, the \emph{capacity constraint} associated with link $(k,l)\in E$ at time~$t$ is given by
\begin{equation} \label{eq.capc}
\textstyle\sum_{n=1}^{N-1} z_{n,t}d_n f_{n,t}^{(k,l)} \leq \min\{c_t^{(k,l)},\hat{c}_{t+1}^{(k,l)}\} \; . 
\end{equation}
\begin{definition}[Feasibility]
The set of routing and admission control decisions at time~$t$, namely $\{f_{n,t}^{(k,l)},z_{n,t}\}$, $\forall n \in V$, $\forall (k,l)$ $\in E$, is \emph{feasible} when it satisfies the flow constraints in \eqref{eq.flow_1}-\eqref{eq.flow_3}, the flow conservation in \eqref{eq.flow} and the capacity constraints in \eqref{eq.capc}.
\end{definition}
\begin{definition}[Max-Min Fairness]\label{def.MaxMin_dyn_routes}
The feasible set $\{f_{n,t}^{(k,l)}$, $z_{n,t}\}$ at time-step $t$ has admission rates $z_{n,t}$ that are max-min fair if, in order to \emph{maintain feasibility}, an increase of any $z_{n,t}$ necessarily results in the decrease of $z_{m,t}$ of another source~$m$ for which $z_{m,t}\leq z_{n,t}$.
\end{definition}

\emph{The goal of the PNR framework} is to dynamically optimize routing and admission control decisions over time, taking into account future predicted network conditions, aiming to maximize the cumulative sum of admission rates $\sum_{t=1}^T\sum_{n=1}^{N-1}z_{n,t}$, while ensuring that, \emph{in each and every time-step}~$t$, the selected feasible set $\{f_{n,t}^{(k,l)},z_{n,t}\}$ is max-min fair and can be implemented without inducing transient congestion. Recall that transient congestion can cause increased delay which can severely affect time-sensitive applications. This challenging optimization problem and its computational complexity are addressed in Sec.~\ref{sec.NRalgorithm}.

\textbf{Real-World Network and Dataset}. 
Consider the backhaul network in Fig.~\ref{CMLSweden}(a) composed of $17$ wireless links whose lengths vary from $0.6$ to $5.9$\thinspace{km} and that operate between $18$ and $40$\thinspace{GHz}. The directed graph $G=(V,E)$ with $N=13$ nodes in Fig.~\ref{CMLSweden}(b) is generated by assuming that \emph{link endpoints} in Fig.~\ref{CMLSweden}(a) that are in close proximity (up to $300$\thinspace{m} apart) are connected by fiber which is not capacity-limited. Under this assumption\footnote{Notice that other assumptions could have been made but they should not affect the generality of the results.}, a node in $G=(V,E)$ represents one or more neighboring link endpoints in Fig.~\ref{CMLSweden}(a). 

The backhaul network in Sweden contains a centralized data collection system (described in detail in \cite{bao31brief}) that periodically gathers measurements from each link $(k,l)\in E$ in intervals of $\Delta=10$ seconds. Each measurement in time-step $t$ includes the transmitted and received signal levels (in dB) represented by $P_{Tx,t}^{(k,l)}$ and $P_{Rx,t}^{(k,l)}$, respectively. 
According to \cite{bao31brief}, the extra load associated with the transmission of measurements via the backhaul network is insignificant. 

In this paper, we consider that, in each time-step $t$, the following events occur: (i) the data collection system shares the latest measurements with the centralized PNR framework; (ii) the AP mechanism predicts the future attenuation levels $x^{(k,l)}_t = P_{Tx,t}^{(k,l)} - P_{Rx,t}^{(k,l)}$ of every link and the MSNR algorithm generates new routing $f_{n,t}^{(k,l)}$ and admission control $z_{n,t}$ decisions; and (iii) the SDN controller implements the new network configuration by propagating $\{f_{n,t}^{(k,l)},z_{n,t}\}$ to the corresponding base-stations. We assume that both the transmission of measurements $\{P_{Tx,t}^{(k,l)},P_{Rx,t}^{(k,l)}\}$ and the propagation of routes and admission control updates $\{f_{n,t}^{(k,l)},z_{n,t}\}$ utilize a negligible amount of resources from the backhaul network. It is important to emphasize that these control packets are transmitted at most once in every $10$ seconds.

To train, validate, and evaluate the PNR framework we use the directed graph $G=(V,E)$ together with a dataset containing $2,295,000$ measurements (i.e., $135,000$ per link) and a \emph{train-validation-test split of 80-10-10}. 
The \emph{test} data utilized to evaluate the PNR framework in Sec.~\ref{sec.Evaluation} is composed of three sequences of measurements, each containing a period of rain: \emph{Test Seq.~I} with $87,890$ measurements collected over a period of $14.3$ hours on 2015-06-02, \emph{Test Seq.~II} with $11,900$ measurements collected over a period of $1.5$ hours on 2015-05-19, and \emph{Test Seq.~III} with $94,690$ measurements collected over a period of $15.5$ hours on 2015-06-17.

\section{Attenuation Prediction Mechanism}\label{sec.APmechanism}
In this section, we present the AP mechanism which predicts future link attenuation levels based on historical data, capturing both \emph{time} and \emph{spatial} correlation that are typical of weather-effects. 
Next, we describe the encoder-decoder LSTM model and the training process. In Sec.~\ref{sec.evalAP}, we compare the performance of the AP mechanism with two benchmark time series prediction methods.


\subsection{Encoder-Decoder LSTM Model}\label{sec.LSTM}
The encoder-decoder LSTM model is a Recurrent Neural Network designed to address sequence-to-sequence prediction problems such as machine translation, natural language generation, and speech recognition~\cite{ehsan2018rationalization, xie2019convolutional,graves2014towards}. The model is composed of two main parts: the \emph{encoder}, which maps the input sequence into a state vector and the \emph{decoder}, which maps the state vector into a sequence of predictions.

\begin{figure}[t] 
\centering
\subfloat[]
{\includegraphics[width=0.3\columnwidth]{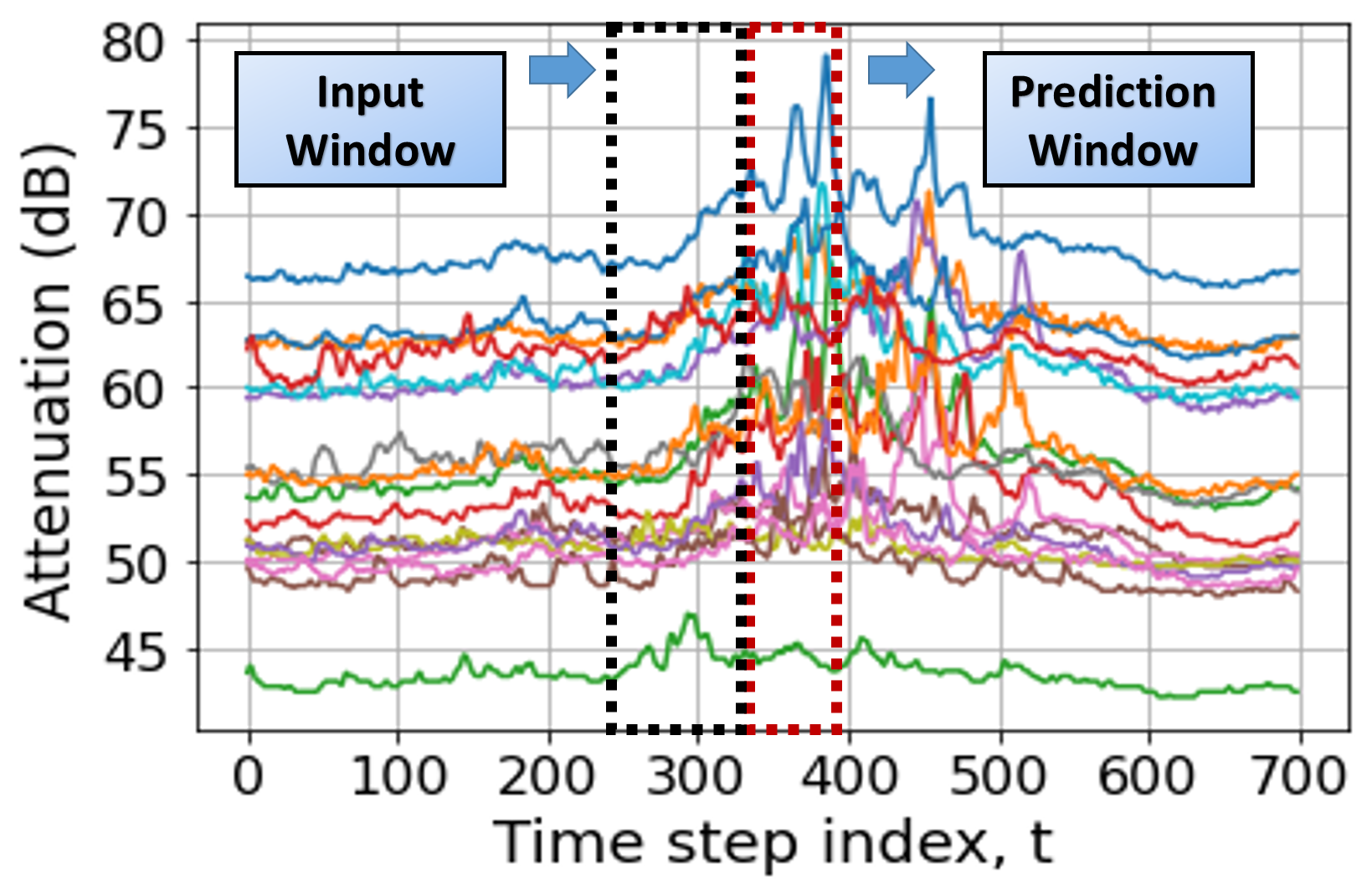}}
\hspace{0.5cm}
\subfloat[]
{\includegraphics[width=0.3\columnwidth]{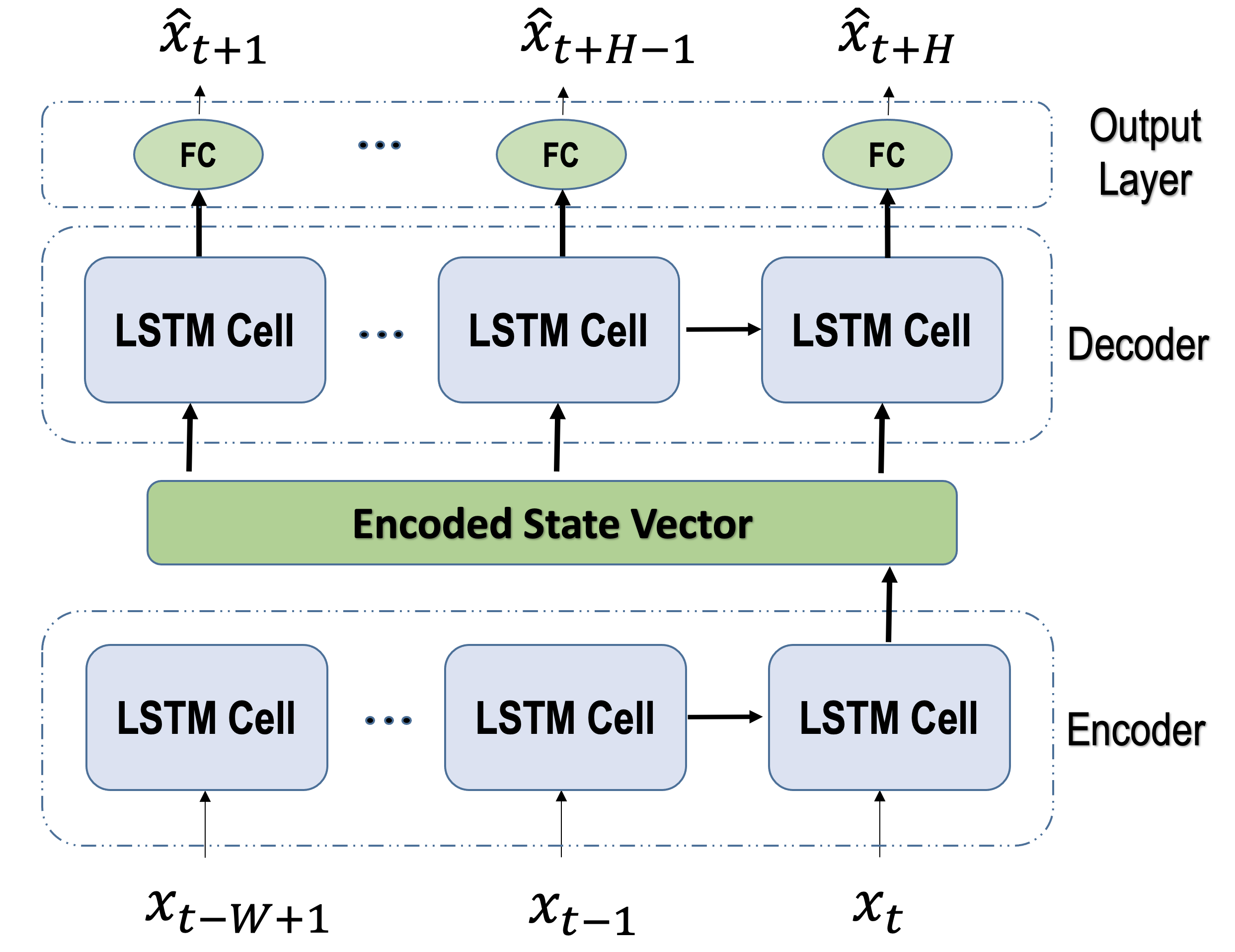}}
\caption{(a) Measured attenuation for all the 17 links in the network in Fig.~\ref{CMLSweden} with time-steps $t$ separated by $10$ seconds (adding up to 1.9 hours) together with an illustration of the sliding-window method with \emph{input window} size of $W$ time-steps and \emph{prediction window} size of $H$ time-steps. The different baseline attenuation levels are due to the different characteristics of the links in terms of distance between base-stations and transmission frequency. An increased attenuation due to rain is observed in the interval between $t=300$ and $600$ time-steps. (b) Encoder-decoder LSTM model that employs the last $W$ measurements $\{\boldsymbol{x_{t-W+1}} ,\boldsymbol{x_{t-W+2}}, \ldots , \boldsymbol{x_t}\}$ from every link in the network (i.e., the input window) to predict the future $H$ attenuation levels $\{\boldsymbol{\hat{x}_{t+1}}$,$\boldsymbol{\hat{x}_{t+2}}, \ldots$, $\boldsymbol{\hat{x}_{t+H}}\}$ in each link (i.e., the prediction window).}\label{fig.APmechanism}
\vspace{-0.5cm}
\end{figure}

The AP mechanism employs the sliding-window method and the encoder-decoder LSTM model illustrated in Figs.~\ref{fig.APmechanism}(a) and \ref{fig.APmechanism}(b), respectively, to predict the next $H$ attenuation levels based on the previous $W$ measurements. In particular, let $x^{(k,l)}_t = P_{Tx,t}^{(k,l)} - P_{Rx,t}^{(k,l)}$ be the \emph{attenuation measurement} for link $(k,l) \in E$ at time~$t$, and let $\boldsymbol{x_t} = (x_t^{(k,l)})$ and $\boldsymbol{\hat{x}_{t+h}} = (\hat{x}_{t+h}^{(k,l)})$ be the vector of \emph{attenuation measurements} and the vector of $h$\emph{-steps-ahead attenuation predictions} for all links at time~$t$, respectively. In each time-step~$t$, the encoder-decoder LSTM model employs the sequence of measurements in the \emph{input window} $\{\boldsymbol{x_{t-W+1}} ,\boldsymbol{x_{t-W+2}}, \ldots , \boldsymbol{x_t}\}$ to predict the sequence of attenuation levels in the \emph{prediction window} $\{\boldsymbol{\hat{x}_{t+1}}$,$\boldsymbol{\hat{x}_{t+2}}, \ldots$, $\boldsymbol{\hat{x}_{t+H}}\}$. Notice that the measurements contained in the \emph{input window} allow the encoder-decoder LSTM model to capture the spatio-temporal correlation that is typical of weather-induced attenuation. We employ an input window size of $W=12$ and a prediction window size of $H=5$, which corresponds to $120$ seconds and $50$ seconds, respectively.

We train the encoder-decoder LSTM model to minimize the prediction error. In particular, consider a dataset with a sequence of attenuation measurements in the interval $t\in\{1,\ldots,T\}$.
The encoder and decoder are jointly trained to minimize the objective function:
\begin{equation}
\mathcal{L}(\Theta) = \textstyle\sum_{t=1}^{T-H} \sum_{h=1}^{H} \left\Vert \;\boldsymbol{x_{t+h}} -\boldsymbol{\hat{x}_{t+h}}\; \right\Vert^2 \; 
\end{equation}
where 
$\left\Vert \cdot \right\Vert$ represents the Euclidean norm, and $\Theta$ represents the parameters of the encoder-decoder LSTM model, i.e., weights and biases. We implement the encoder and the decoder LSTM with one hidden layer containing 128 units.
We use the dataset collected from the backhaul network to train, tune, and evaluate the AP mechanism. We train the AP mechanism using Backpropagation Through Time \cite{werbos1990backpropagation} and Adaptive Moment Estimation (Adam) \cite{kingma2014adam} with a batch size of $150$. The prediction accuracy of the AP mechanism is evaluated in Sec.~\ref{sec.evalAP}.

\section{Multi-Step Network Reconfiguration Algorithm}\label{sec.NRalgorithm}
SDN enables the design of algorithms that dynamically reconfigure the entire network. Building on that, in this section, we develop the MSNR algorithm, which leverages information about links' future conditions to compute the sequence of current and future routing and admission control decisions that attempt to maximize network utilization, while achieving max-min fairness (in every time-step $t$) among the base-stations sharing the network and preventing transient congestion that may be caused by re-routing. Hereafter, we denote this sequence of routing and admission control decisions as the \emph{optimal sequence of network configurations}. 


The problem of finding the optimal sequence of network configurations is a generalization of the well-known Maximum Concurrent Flow (MCF) problem \cite{allalouf2004maximum,ShahrokhiM88} for the more challenging setting where: (i) a sequence of predictions of future network conditions are available and (ii) transient congestion due to re-routing is taken into account. The MSNR algorithm employs MPC to address this generalized MCF optimization problem. In particular, in each time-step~$t$, the MSNR algorithm uses its knowledge of future (predicted) network conditions to evaluate and compare the performance of different congestion-free sequences of network configurations $\{f_{n,t+h}^{(k,l)},z_{n,t+h}\}$, $\forall n \in V$, $\forall (k,l)$ $\in E$, $\forall h\in\{0,1,\ldots,H-1\}$ and then it selects the max-min fair sequence that maximizes the cumulative sum of admission rates $\sum_{h=0}^{H-1}\sum_{n=1}^{N-1}z_{n,t+h}$. The SDN controller implements the first configuration in the selected sequence, i.e., the configuration $\{f_{n,t}^{(k,l)},z_{n,t}\}$ associated with the current time~$t$. This iterative process allows the SDN controller to account for future predicted network conditions when optimizing the current network configuration.


An important challenge associated with the MSNR algorithm is computational complexity. A naive implementation of the MSNR algorithm computes and compares the performance of all possible sequences of network configurations within the prediction window $\{t,\ldots,t+H\}$. The number of such sequences grows exponentially with $H$, as we will discuss in Sec.~\ref{sec.Multple}, which could render the MSNR algorithm impractical. To overcome this challenge, we develop a principled implementation of the MSNR algorithm which employs the structure of the optimization problem to recursively explore the space of all possible sequences of network configurations. This recursive method reduces the complexity from exponential $O(2^H)$ to polynomial $O(H^4)$.

Prior to introducing the MSNR algorithm, we describe: (i) the adaptive modulation mechanism in \cite{bao2015field}, which is a Physical layer mechanism employed by the backhaul network in Sweden to maximize link capacity over time; and (ii) the SWAN mechanism developed in \cite{Hong2013Achieving}, which is a Network layer mechanisms that eliminates transient congestion that may be caused by re-routing. The MSNR algorithm builds upon  both these existing solutions to address the generalized MCF problem, enabling the optimization of routing and admission control decisions over time in a setting where predictions of future network conditions are available.

\subsection{Adaptive Modulation Mechanism}\label{sec.AMM}
Three parameters that can be dynamically adjusted to compensate for high attenuation levels in microwave and mmWave links are: the transmission power, the coding rate, and the modulation scheme. The dataset utilized in this paper was collected for a backhaul network that uses radios similar to the ones described in ~\cite{bao2015field, bao31brief} which: 
(i) employ a constant transmit power $P_{Tx,t}^{(k,l)}$ and a constant coding rate over time; (ii) use Quadrature Amplitude Modulation (QAM) with \emph{adaptive} constellation size $\modulation$; and (iii) use a fixed channel bandwidth of $28$\thinspace{MHz} that achieves a capacity of $45$\thinspace{Mbps} when $\modulation=4$. Recall that when $\modulation$ is increased by a factor of $k$, the capacity $c_t^{(k,l)}$ increases by a factor of $\log_2{k}$ and the Bit Error Rate (BER) decreases according to \cite[Eq.\thinspace{(18)}]{BER}. 

The adaptive modulation (AM) mechanism adjusts the constellation size $\modulation$ over time, aiming to maximize link capacity $c_t^{(k,l)}$ while keeping the BER above a given threshold. For complying with the description of the radios in \cite[Sec.~II.B]{bao2015field}, hereafter in this paper, we consider a wireless x-haul network that employs the AM mechanism with hysteresis represented in Table~\ref{tab.adaptiveModulation}. In particular, we consider that every link $(k,l) \in E$ uses radios that adapt their constellation size $\modulation$ at each time-step~$t$ based on Table~\ref{tab.adaptiveModulation} and on their measured received signal level $P_{Rx,t}^{(k,l)}$. The \emph{limit up} in Table~\ref{tab.adaptiveModulation} represents the received signal level in which the adopted $\modulation$ should increase. The \emph{limit down} represents the received signal level in which the adopted $\modulation$ should decrease to keep the BER above the set threshold. 
Notice that Table~\ref{tab.adaptiveModulation} represents a mapping from the evolution of the received signal levels $P_{Rx,t}^{(k,l)}$ over time to the evolution of the link capacities $c_t^{(k,l)}$ over time.

\begin{table}[t]
\caption{Parameters associated with the adaptive modulation mechanism with hysteresis for a BER threshold of $10^{-9}$.}\label{tab.adaptiveModulation}
\begin{center}
\begin{tabular}{cccc} \toprule
$\modulation$ & Bitrate (Mbps) & Limit up (dBm) & Limit down (dBm)\\
\midrule
4       &45     &-72    &N/A    \\
16      &90     &-66    &-74    \\
64      &135    &-62.5  &-68    \\
128     &157    &-61    &-64    \\
256     &180    &-57    &-62    \\
512     &202.5  &-53    &-58    \\
1024    &225    &N/A    &-54    \\
\bottomrule
\end{tabular}
\end{center}
\end{table}

\subsection{The Cost of Re-routing}\label{sec.Cost}


One possible approach to dynamically optimizing the network configuration 
\emph{without resorting to predictions of links' future conditions} is for the SDN controller to carry out, in each time~$t$, the following procedure: (i) gather information about the current link capacities $c_t^{(k,l)}$; (ii) employ existing solutions to the MCF optimization problem (e.g., \cite{allalouf2004maximum,ShahrokhiM88}) to find the configuration $\{f_{n,t}^{(k,l)},z_{n,t}\}$ that maximizes the \emph{current} network utilization; and (iii) implement the new routing decisions $f_{n,t}^{(k,l)}$ and admission rates $z_{n,t}$ by sending control packets to the base-stations in the x-haul network. Upon reception of these control packets, the base-stations add/remove entries from their routing tables and adjust their network slice admission and provisioning accordingly. Two important drawbacks of this approach are the delay to recover from performance drops, which is characteristic of reactive reconfiguration mechanisms, and that it does not take into account the transient congestion that may be caused by re-routing. Both drawbacks may severely affect time-sensitive traffic. The MSNR algorithm proposed in Sec.~\ref{sec.Multple} addresses both drawbacks. In this section, we discuss the negative effects that re-routing may have on the network performance.

To update routes from $f_{n,t-1}^{(k,l)}$ to $f_{n,t}^{(k,l)}$, the SDN controller may have to send control packets to multiple base-stations. 
Due to communication and processing delays, some base-stations may apply the new routes $f_{n,t}^{(k,l)}$ while others still employ old routes $f_{n,t-1}^{(k,l)}$, which may cause significant transient congestion and over-utilization of communication links, namely violation of the capacity constraints in \eqref{eq.capc}. Depending on the duration and magnitude of the congestion, data packets may be severely delayed or even lost. 
In this case, the re-routing process is clearly imposing a performance cost that should be taken into account when the SDN controller decides whether or not to re-route. 


In order to reduce the transient congestion associated with re-routing, a common approach (e.g., \cite{Yaghoubi2018,Hong2013Achieving, zUpdate2013}) is to subdivide the re-routing  process into multiple stages. In each stage, the SDN controller updates a small number of base-stations, instead of updating all of them at the same time. Each stage is designed to generate zero (or little) transient congestion and the complete sequence of stages is designed to lead to the desired final routing configuration. An important constraint is that the time for completing the re-routing process should be shorter than the interval between two consecutive time-steps, e.g., $t$ and $t+1$, which in this paper is of $\Delta=10$ seconds. In~\cite{Yaghoubi2018,Hong2013Achieving,zUpdate2013}, the authors propose different route implementation systems that attempt to minimize the transient congestion. 
Yet, these route implementation systems can only guarantee that re-routing is performed with zero congestion when a portion of the network capacity is vacant before the update. Naturally, when all links are fully utilized, the first update to take effect will always congest at least one link.

In this paper, we consider an SDN controller that implements any given set of new routes $f_{n,t}^{(k,l)}$ by employing the \emph{SWAN mechanism} developed in \cite{Hong2013Achieving}. The SWAN mechanism leverages \emph{scratch capacity} in \emph{every link} to perform congestion-free re-routing. In particular, the authors of \cite{Hong2013Achieving} show that SWAN can update routes, i.e., change from $f_{n,t-1}^{(k,l)}$ to any given $f_{n,t}^{(k,l)}$, with zero transient congestion in \emph{at most} $\lceil 1/\scratch_t \rceil - 1$ stages, where $\scratch_t\in(0,1]$ represents the \emph{scratch capacity of the network} at time~$t$. Formally, $\scratch_t$ is given by
\begin{equation}\label{eq.scratch}
\scratch_t =\mbox{ argmax} \left\{\scratch\in(0,1] \;\middle|\; \sum_{n=1}^{N-1} z_{n,t-1}d_n f_{n,t-1}^{(k,l)} \leq (1-\scratch) c_t^{(k,l)}, \forall (k,l)\in E\right\}.
\end{equation}
For details on how the SWAN route implementation mechanism works, we refer the reader to \cite{Hong2013Achieving}. 
Notice that when the network has no \emph{scratch capacity}, i.e., $\scratch_t \rightarrow 0$, the SWAN mechanism needs $\lceil 1/\scratch_t \rceil - 1 \rightarrow \infty$ stages to complete a single congestion-free re-routing process. To limit the re-routing time, we impose a lower bound of $\scratch_{min}=0.05$ on the \emph{scratch capacity}, $\scratch_t$, needed for a re-route. Hereafter in this paper, we assume that the SDN controller is allowed to re-route at time~$t$ if and only if $\scratch_t \geq \scratch_{min}=0.05$.

The SDN controller employs the MSNR algorithm to compute the optimal sequence of network configurations over time and, when necessary, it employs the SWAN mechanism to implement new routes. In particular, in each time-step~$t$, given the prior routing and admission control decisions, $\{f_{n,t-1}^{(k,l)},$ $z_{n,t-1}\}$, the SDN controller employs \eqref{eq.scratch} to calculate the current scratch capacity $\scratch_t$. If $\scratch_t \geq \scratch_{min}$, the SDN controller employs the MSNR algorithm to compute the optimal sequence of network configurations and then it employs SWAN to implement the optimal configuration $\{f_{n,t}^{(k,l)},$ $z_{n,t}\}$ at the current time~$t$. Alternatively, if $\scratch_t < \scratch_{min}$, the SDN controller is not allowed to re-route at time~$t$, but it can still optimize the admission rates $z_{n,t}$. In this case, the SDN controller employs the MSNR algorithm \emph{with fixed routing parameters} $f_{n,t}^{(k,l)} = f_{n,t-1}^{(k,l)}$ to compute the optimal sequence of network configurations and then it implements the optimal admission rates $z_{n,t}$ at time~$t$. It is easy to see that admission rates can be updated from $z_{n,t-1}$ to $z_{n,t}$ with zero transient congestion in at most two stages, irrespective of the value of $\scratch_t$. In the first stage, the SDN controller updates all base-stations in which $z_{n,t-1}>z_{n,t}$ and, in the last stage, the SDN controller updates all base-stations in which $z_{n,t-1}<z_{n,t}$. 

The routing and admission control decisions \emph{at time}~$t$ determine the scratch capacity $\scratch_{t+1}$ \emph{at time}~$t+1$, which determines whether or not the SDN controller will be allowed to re-route at time~$t+1$. Hence, if the SDN controller plans to re-route at time~$t+1$, it should select a network configuration $\{f_{n,t}^{(k,l)},$ $z_{n,t}\}$ that will lead to $\scratch_{t+1} \geq \scratch_{min}$. This can be achieved by employing, at time~$t$, the following capacity constraint for every link $(k,l)\in E$
\begin{equation} \label{eq.capc_scratch}
\textstyle\sum_{n=1}^{N-1} z_{n,t}d_n f_{n,t}^{(k,l)} \leq \min\{c_t^{(k,l)},(1-\scratch_{min})\hat{c}_{t+1}^{(k,l)}\} \; .
\end{equation}
Alternatively, if the SDN controller plans to keep the same routes in the next time-step, i.e., $f_{n,t+1}^{(k,l)} = f_{n,t}^{(k,l)}$, it should attempt to fully utilize the links, leaving no scratch capacity. This can be achieved by employing the capacity constraints in \eqref{eq.capc}. Intuitively, this means that, in order to re-route in the next time-step~$t+1$, the SDN controller may need to reduce the admission rates $z_{n,t}$ in the current time-step~$t$. \emph{This potential reduction of} $z_{n,t}$  \emph{represents the cost of re-routing}, as illustrated in the following example. 

\begin{figure}[t]
\centering
\includegraphics[width=0.55\columnwidth]{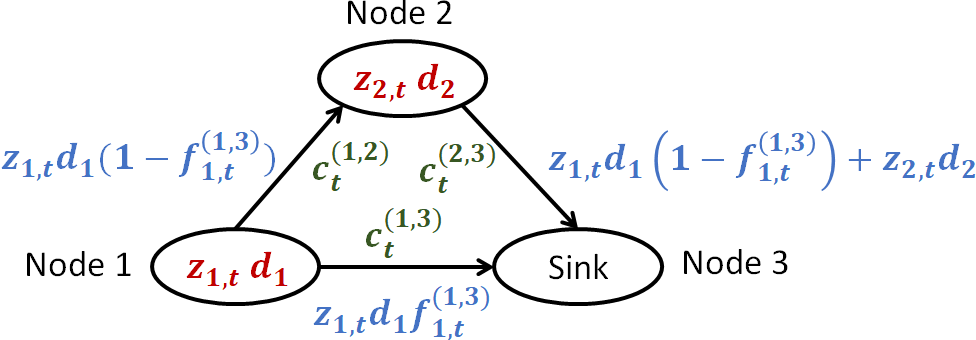}
\vspace{-0.2cm}
\caption{Illustration of a network with $N=3$ nodes (two commodities and a destination) and three links. The admitted demands $z_{n,t}d_n$ at time-step~$t$ are shown within the corresponding nodes. The total flows and capacities at time-step~$t$ are shown next to the corresponding links.} \label{fig.ToyNetwork}
\vspace{-0.4cm}
\end{figure}

\textbf{Example:} consider the network in Fig.~\ref{fig.ToyNetwork} with $N=3$ nodes, three links $\{(1,2),(2,3),(1,3)\}$, and fixed demands $d_1=1$ and $d_2=0.5$. Assume that this network has capacities $c_{t}^{(k,l)}=0.5$ for all links and predicted capacities $\hat c_{t+h}^{(k,l)}=0.5$ for all links and prediction horizons $h$. Moreover, assume that $\scratch_{t}\geq\scratch_{min}=0.05$, meaning that the SDN controller is allowed to re-route at the current decision time~$t$. 

\emph{Plan to not re-route.} If the SDN controller plans to keep the same routes in future time-steps, then it adopts the capacity constraints in \eqref{eq.capc}. In this toy example, it is easy to see that the corresponding max-min fair admission rates are $z_{1,t}=z_{2,t}=2/3$.  
Notice that there exists feasible configurations $\{f_{n,t}^{(k,l)},z_{n,t}\}$ with higher sum $\sum_{n=1}^{2}z_{n,t}$, but their admission rates are not max-min fair. An example of such unfair feasible admission rates are $z_{1,t}=0.5$, $z_{2,t}=1$.  

\emph{Plan to re-route}. Alternatively, if the SDN controller plans to re-route in the next time-step, then it adopts the capacity constraints in \eqref{eq.capc_scratch} with $\scratch_{min}=0.05$. It is easy to see that the corresponding max-min fair admission rates are $z_{1,t}=z_{2,t}=2/3*(1-0.05)$.

Two important observations are: (i) Planning to re-route at time~$t+1$ does not guarantee that the SDN controller will be able to re-route at time~$t+1$. In particular, if the capacity prediction is inaccurate and (by chance) $\hat c_{t+1}^{(k,l)} > c_{t+1}^{(k,l)}$, the SDN controller may not have enough scratch capacity at time~$t+1$ to re-route. (ii) Planning to re-route at time~$t+1$, can only hurt the network performance at the current time~$t$ due to the provision of the scratch capacity, as illustrated in the example. The potential benefits of planning to re-route at time~$t+1$ can only be assessed by computing the performance of the network at future time-steps. 

\subsection{Optimal Sequence of Network Configurations}\label{sec.Multple}

In this section, we develop the MSNR algorithm which leverages information about current and future predicted link capacities $\{c_t^{(k,l)},$ $\hat{c}_{t+1}^{(k,l)},$ $\ldots,$ $\hat{c}_{t+H}^{(k,l)} \}$ to dynamically optimize routing and admission control decisions aiming to maximize the cumulative sum of admission rates $\sum_{t=1}^T\sum_{n=1}^{N-1}z_{n,t}$, while ensuring that, \emph{in every time-step} $t$, the selected feasible set $\{f_{n,t}^{(k,l)},z_{n,t}\}$ is max-min fair and can be implemented by the SDN controller without inducing transient congestion. The MSNR algorithm addresses a generalization of the MCF problem \cite{allalouf2004maximum,ShahrokhiM88} for the more challenging setting where: (i) a sequence of predictions of future network conditions are available and (ii) transient congestion due to re-routing is taken into account.


Prior to describing the MSNR algorithm, we introduce the concept of a re-routing plan. For a given time~$t$ and a prediction window size $H$, let $\route_{t,h}$ be an indicator function that is equal to $1$, if the plan is to re-route in time-step~$t+h$, $\forall h\in\{0,1,\ldots,H\}$, and $\route_{t,h}=0$, otherwise. The \emph{re-routing plan} at time~$t$ is given by the vector $\mathbf{\route_t}=(\route_{t,0},\route_{t,1},\ldots,\route_{t,H})$. Notice that if $\scratch_t<\scratch_{min}$, then $\route_{t,0}=0$ and if $\scratch_t\geq\scratch_{min}$, then $\route_{t,0}\in\{0,1\}$. Next, we use an example to show how the re-routing plan $\mathbf{\route_t}$ can be utilized to separate the problem of finding the optimal sequence of network configurations into simpler sub-problems. 

\textbf{Example:} consider a network with a prediction window size of $H=5$ and a 
plan $\mathbf{\route_t}=(0,1,0,0,1,0)$ to re-route only at times $t+1$ and $t+4$. The network parameters associated with this particular plan $\mathbf{\route_t}$ are displayed in Table~\ref{tab.parameters}. The capacity constraints in time-step~$t+h$ depend on whether the plan is to re-route in the next time-step~$t+h+1$ or not, according to the following expression
\begin{equation} \label{eq.capc_scratch_joint}
\textstyle\sum_{n=1}^{N-1} z_{n,t+h}d_n f_{n,t+h}^{(k,l)}  \leq \min\{\hat{c}_{t+h}^{(k,l)},(1-\scratch_{min}\route_{t,h+1})\hat{c}_{t+h+1}^{(k,l)}\}, \forall (k,l)\in E \; .
\end{equation}
Equation \eqref{eq.capc_scratch_joint} is a generalization of \eqref{eq.capc} and \eqref{eq.capc_scratch}. The second column of Table~\ref{tab.parameters} represents the RHS of the capacity constraint in \eqref{eq.capc_scratch_joint}. Notice from Table~\ref{tab.parameters} that $f_{n,t+h}^{(k,l)}$ can be updated only at the re-routing times $t+1$ and $t+4$ while $z_{n,t+h}$ can be updated at every time-step. Hence, the routing decisions at time~$t+1$, namely $f_{n,t+1}^{(k,l)}$, affect not only $z_{n,t+1}$, but also $z_{n,t+2}$ and $z_{n,t+3}$. It follows that the optimization problem associated with $\mathbf{\route_t}=(0,1,0,0,1,0)$ can be subdivided at the re-routing times, resulting in three simpler sub-problems, each of which jointly optimizes: (i) $f_{n,t}^{(k,l)}$ and $z_{n,t}$; (ii) $f_{n,t+1}^{(k,l)}$, $z_{n,t+1}$, $z_{n,t+2}$, and $z_{n,t+3}$; and (iii) $f_{n,t+4}^{(k,l)}$ and $z_{n,t+4}$. 

\begin{table}[t]
\caption{Evolution of network parameters associated with the re-routing plan $\mathbf{\route_t}=(0,1,0,0,1,0)$.}\label{tab.parameters}
\vspace{-0.4cm}
\begin{center}
\begin{tabular}{cccc} \toprule
plan & capacity constraints in \eqref{eq.capc_scratch_joint} & admis. & routing \\
\midrule
$\route_{t,0}=0$ & $\min\{c_t^{(k,l)},(1-\scratch_{min})\hat{c}_{t+1}^{(k,l)}\}$ & $z_{n,t}$ & $f_{n,t-1}^{(k,l)}$ \\
$\route_{t,1}=1$ & $\min\{\hat{c}_{t+1}^{(k,l)},\hat{c}_{t+2}^{(k,l)}\}$ & $z_{n,t+1}$ & $f_{n,t+1}^{(k,l)}$ \\
$\route_{t,2}=0$ & $\min\{\hat{c}_{t+2}^{(k,l)},\hat{c}_{t+3}^{(k,l)}\}$ & $z_{n,t+2}$ & $f_{n,t+1}^{(k,l)}$ \\
$\route_{t,3}=0$ & $\min\{\hat{c}_{t+3}^{(k,l)},(1-\scratch_{min})\hat{c}_{t+4}^{(k,l)}\}$ & $z_{n,t+3}$ & $f_{n,t+1}^{(k,l)}$ \\
$\route_{t,4}=1$ & $\min\{\hat{c}_{t+4}^{(k,l)},\hat{c}_{t+5}^{(k,l)}\}$ & $z_{n,t+4}$ & $f_{n,t+4}^{(k,l)}$ \\
$\route_{t,5}=0$ & $\hat{c}_{t+6}^{(k,l)}$ is unknown & N/A & N/A \\
\bottomrule
\end{tabular}
\end{center}
\vspace{-0.4cm}
\end{table}

In general, the optimization problem associated with any re-routing plan $\mathbf{\route_t}$ can be subdivided at the re-routing times (i.e., times $t+h$ in which $\route_{t,h}=1$) \emph{without loss of optimality}. Let $\{t+h_1,$ $t+h_1+1,$ $\ldots,$ $t+h_2\}$ represent a subdivision of a re-routing plan $\mathbf{\route_t}$. The Generalized-MCF (G-MCF) algorithm described in Algorithm~\ref{alg.Generalized_MCFalgorithm} jointly optimizes the routing decisions $f_{n,t+h_1}^{(k,l)}, \forall n, \forall (k,l)$  at the initial time $t+h_1$ and the admission rates $z_{n,t+h}, \forall n, \forall h\in\{h_1,\ldots,h_2\}$. To address this joint optimization, G-MCF solves a sequence of MCF problems with increasing admission rates $z_{n,t+h}$ until all commodities in the network become saturated.  In particular, let $k$ be the iteration index of the algorithm, let $\unsaturated$ be the set of unsaturated commodities at the beginning of iteration~$k$, let $(n,h)$ be the tuple that represents the index of the $(N-1)(h_2-h_1+1)$ different commodities in this subdivision of $\mathbf{\route_t}$, and let $z_{(n,h)}^S$ be the admission rate $z_{n,t+h}$ that saturates commodity~$(n,h)$. Initially, we have $\unsaturated=\{(n,h)\},\forall n,h$. In each iteration~$k$, the algorithm solves the MCF problem associated with the \emph{unsaturated commodities}, i.e., it assigns $z_{n,t+h}\leftarrow \bar z, \forall (n,h)\in\unsaturated$, and $z_{n,t+h}\leftarrow z_{(n,h)}^S, \forall (n,h)\notin\unsaturated$, and finds the feasible configuration (Output~1) with maximum value of $\bar z \in[0,1]$ which we denote by $\bar z^*$. Then, the algorithm identifies the commodities that become saturated\footnote{Identifying the commodities that become saturated in iteration~$k$ is not straightforward. The authors in \cite{allalouf2004maximum} developed a saturation test which we adapt to our network setting in lines~10 - 23 of Algorithm~\ref{alg.Generalized_MCFalgorithm}.} in the current iteration, stores their saturation values $z_{(n,h)}^S \leftarrow \bar z^*$, updates the set $\unsaturated$ accordingly, and proceeds to the next iteration $k+1$. The algorithm terminates when $\unsaturated=\emptyset$ and the admission rates $z_{(n,h)}^S$ that saturate every commodity in the network have been determined. The G-MCF algorithm finds the optimal routing and admission control decisions within a  subdivision $\{t+h_1,$ $t+h_1+1,$ $\ldots,$ $t+h_2\}$ of the re-routing plan $\mathbf{\route_t}$. To compute the optimal routing and admission control decisions associated with an entire re-routing plan $\mathbf{\route_t}=$ $(\route_{t,0},$ $\route_{t,1},$ $\ldots,\route_{t,H})$, the G-MCF algorithm is utilized in each of its subdivisions.

\begin{algorithm}
\small
\SetAlgoLined 
$\%$ Let $\{t+h_1,\ldots,t+h_2\}$ be the subdivision of the re-routing plan $\mathbf{\route_t}$ under consideration and let $\unsaturated$ be the set of unsaturated commodities at the beginning of iteration $k$\;
Initialization: $\unsaturated=\{(n,h)\},$ $\forall n\in\{1,2,\ldots,N-1\},$ $\forall h\in\{h_1,\ldots,h_2\}$ 
and $k=0$\;
 \While{$\unsaturated\neq\emptyset$}
 {
 $\%$ Find $\bar z$ that solves the joint optimization\;
 \For{$n\in\{1,\ldots,N-1\}$ and $h\in\{h_1,\ldots,h_2\}$}{
 \textbf{if} $(n,h)\in\unsaturated$ \textbf{then} $z_{n,t+h}\leftarrow \bar z$\;
 \textbf{else} $z_{n,t+h}\leftarrow z_{(n,h)}^S$\; }
 Solve: $\max \bar z$, s.t. $\bar z\in[0,1]$, and \eqref{eq.flow_1}-\eqref{eq.flow}, and capacity constraints in \eqref{eq.capc_scratch_joint} for $h\in\{h_1,\ldots,h_2\}$\;
 Output $1$: values of $\bar z^*$ and $f_{n,t+h}^{(k,l)}$\;
 $\%$ Identify the new saturated commodities $(n,h)$\;
 Determine the set $\disconnected$ of disconnected commodities in the residual graph associated with Output $1$\;
 $Saturation\;Flag =\emptyset$\;
 \For{$(n,h)\in\disconnected$}{
 $\%$ Find $\bar z_{(n,h)}$ that solves the joint optimization\;
 Assign: $z_{n,t+h}\leftarrow \bar z_{(n,h)}$\;
 \textbf{for} $(m,j)\in\unsaturated\setminus (n,h)$ \textbf{do} $z_{m,t+j}\leftarrow \bar z^*$\; 
 Solve: $\max \bar z_{(n,h)}$, s.t. $\bar z_{(n,h)}\in[0,1]$, and \eqref{eq.flow_1}-\eqref{eq.flow}, and constraints in \eqref{eq.capc_scratch_joint} for $h\in\{h_1,\ldots,h_2\}$\;
 Output $2$: values of $\bar z^*_{(n,h)}$ and $f_{n,t+h}^{(k,l)}$\;
 \If{$\bar z^*_{(n,h)} = \bar z^*$}{
 $Saturation\:Flag \leftarrow Saturation\:Flag \cup (n,h)$\;
 }
 }
 \For{$(n,h)\in$ Saturation Flag}{
 Assign: $z_{(n,h)}^S\leftarrow\bar z^*$\;
 Assign: $\unsaturated \leftarrow\unsaturated\setminus (n,h)$\;
 }
 $k \leftarrow k + 1$
 }
 $\%$ Find the max-min fair feasible configuration\;
 \For{$n\in\{1,\ldots,N-1\}$ and $h\in\{h_1,\ldots,h_2\}$}{Assign: $z_{n,t+h}\leftarrow z_{(n,h)}^S$\;}
 Obtain: $f_{n,t+h}^{(k,l)}$ that satisfy \eqref{eq.flow_1}-\eqref{eq.flow} and capacity constraints in \eqref{eq.capc_scratch_joint} for $h\in\{h_1,\ldots,h_2\}$\;
 Output $3$: values of $z_{n,t+h}=z_{(n,h)}^S$ and $f_{n,t+h}^{(k,l)}$\;
 \caption{Generalized-MCF (G-MCF) algorithm}\label{alg.Generalized_MCFalgorithm}
\end{algorithm}


%
\textbf{MSNR algorithm.} To find the optimal sequence of network configurations at time-step $t$, the MSNR algorithm selects the plan $\mathbf{\route_t^*}$ with highest cumulative sum of admission rates $\sum_{h=0}^{H-1}\sum_{n=1}^{N-1}z_{n,t+h}$. A naive implementation of the MSNR algorithm computes and compares the performance of the (at least) $2^{H}$ admissible re-routing plans. 
To reduce the computational complexity from exponential $O(2^H)$ to polynomial $O(H^4)$, we propose a principled implementation of the MSNR algorithm based on backward induction which leverages the fact that the optimization problem can be subdivided at the re-routing times without loss of optimality. Specifically, the algorithm separates re-routing plans $\mathbf{\route_t}$ into $H+1$ disjoint sets and then finds the best plan within each set. The first set contains plans that re-route \emph{for the first time} at step~$t+H-1$, the second set contains plans that re-route \emph{for the first time} at step~$t+H-2$, and so on, until the last set which contains a plan that never re-routes. A key observation is that computations for earlier sets can be used to simplify computations for later sets. A description of this computation is provided below.

\emph{First Set.} Consider plans $\mathbf{\route_t}$ that re-route \emph{for the first time} at step~$t+H-1$, i.e., $\mathbf{\route_t}\in\{(0,\ldots,0,1,0)$, $(0,\ldots,0,1,1)\}$. The MSNR algorithm employs the G-MCF algorithm to compute the optimal routing and admission control decisions for these $2$ re-routing plans and then selects the plan $\mathbf{\route_t^{(1)}}$ with highest cumulative sum of admission rates at time $t+H-1$, namely $\sum_{n=1}^{N-1}z_{n,t+H-1}$. 

\emph{Second Set.} Consider plans that re-route for the first time at step~$t+H-2$, i.e., $\mathbf{\route_t}\in\{(0,\ldots,0,1,0,0)$, $(0,\ldots,0,1,0,1)$, $(0,\ldots,0,1,1,0)$, $(0,\ldots,0,1,1,1)\}$. Notice that in the \emph{subset of plans} that re-route both at times $t+H-2$ and $t+H-1$, we know from the First Set  that $\mathbf{\route_t}=\mathbf{\route_t^{(1)}}+(0,\ldots,0,1,0,0)$ has the best performance and, hence, all other plans in this particular subset can be excluded from consideration. The MSNR algorithm computes the optimal routing and admission control decisions for the remaining $3$ re-routing plans and selects the plan $\mathbf{\route_t^{(2)}}$ with highest cumulative sum of admission rates at times $t+H-2$ and $t+H-1$, namely $\sum_{h=H-2}^{H-1}\sum_{n=1}^{N-1}z_{n,t+h}$. 

\emph{Third Set.} Consider plans that re-route for the first time at step~$t+H-3$, i.e., $\mathbf{\route_t}\in\{(0,\ldots,0,1,0,0,0)$, $\ldots$, $(0,\ldots,0,1,1,1,0)$, $(0,\ldots,0,1,1,1,1)\}$. Notice that in the \emph{subset of plans} that re-route both at times $t+H-3$ and $t+H-2$, we know from the Second Set that $\mathbf{\route_t}=\mathbf{\route_t^{(2)}}+(0,\ldots,0,1,0,0,0)$ has the best performance and, hence, all other plans in this particular subset can be excluded from consideration. Similarly, in the \emph{subset of plans} that re-route both at times~$t+H-3$ and~$t+H-1$, but do not re-route at time~$t+H-2$, we know from the First Set that $\mathbf{\route_t}=\mathbf{\route_t^{(1)}}+(0,\ldots,0,1,0,0,0)$ has the best performance and, hence, all other plans in this particular subset can be excluded from consideration. The MSNR algorithm computes the optimal routing and admission control decisions for the remaining $4$ re-routing plans and selects the plan $\mathbf{\route_t^{(3)}}$ with highest cumulative sum of admission rates from times $t+H-3$ to $t+H-1$, namely $\sum_{h=H-3}^{H-1}\sum_{n=1}^{N-1}z_{n,t+h}$. 

\emph{Subsequent Sets.} The MSNR algorithm considers the set of plans that re-route for the first time at steps~$t+H-4$, $t+H-5$, $\ldots$, $t$ and employs an analogous procedure in order to determine the best plans $\mathbf{\route_t^{(4)}}$, $\mathbf{\route_t^{(5)}}$, $\ldots$, $\mathbf{\route_t^{(H)}}$.

\emph{Last Set.} The MSNR algorithm compares the performance of the best plans $\mathbf{\route_t^{(h)}},\forall h\in\{1,2,\cdots,H\}$ with the performance of the never re-route plan $(0,\ldots,0,0)$ and then selects the plan $\mathbf{\route_t^{*}}$ with highest cumulative sum of admission rates $\sum_{h=0}^{H-1}\sum_{n=1}^{N-1}z_{n,t+h}$ in the entire prediction window. The routing and admission control decisions associated with $\mathbf{\route_t^{*}}$ are the optimal sequence of network configurations. 

\begin{remark}[Computational Complexity]\label{rem.complexity}
To find the best plans $\mathbf{\route_t^{(1)}}$, $\mathbf{\route_t^{(2)}}$, $\ldots$, $\mathbf{\route_t^{(H)}}$ in each of the corresponding backward induction steps, the MSNR algorithm computes and compares the performance of $2,3,\ldots,H+1$ re-routing plans, respectively. Then, in the last step of the induction, the MSNR algorithm computes and compares the performance of $H+2$ re-routing plans in order to find the plan $\mathbf{\route_t^{*}}$ and the associated optimal sequence of network configurations $\{f_{n,t+h}^{(k,l)},$ $z_{n,t+h}\}$, $\forall h\in\{0,1,\ldots,H-1\}$, at time $t$. In total, the MSNR algorithm employing backward induction computes the performance of $(H+1)(H+4)/2$ re-routing plans, as opposed to the (at least) $2^{H}$ computations associated with the naive implementation. 
Notice from Algorithm~\ref{alg.Generalized_MCFalgorithm} that to compute the performance of any given re-routing plan $\mathbf{\route_t}$, the G-MCF algorithm solves $O(H^2N^2)$ MCF optimization problems, each of which can be solved in polynomial time \cite{allalouf2004maximum,ShahrokhiM88}. It follows that the MSNR algorithm has polynomial computational complexity which grows as $O(H^4)$.
\end{remark}

\begin{proposition}[Max-Min Fairness of the MSNR algorithm]\label{prop}
The optimal sequence of network configurations $\{f_{n,t+h}^{(k,l)},$ $z_{n,t+h}\}$ given by the MSNR algorithm has admission rates $\{z_{n,t+h}\}_{n=1}^{N-1}$ that are max-min fair in every time-step $t+h$ for any given $h\in\{0,1,\ldots,H-1\}$, irrespective of the network topology $G=(V,E)$, demands $d_n$, and current and predicted link capacities $\{c_t^{(k,l)},$ $\hat{c}_{t+1}^{(k,l)},$ $\ldots,$ $\hat{c}_{t+H}^{(k,l)} \}$. 
\end{proposition}

\begin{IEEEproof}
Proposition~\ref{prop} holds by the design of the MSNR algorithm. 
In the first iteration, Algorithm~\ref{alg.Generalized_MCFalgorithm} finds the \emph{lowest} admission rate $\bar z^*$ that saturates at least one commodity $(n,h)$, assigns $z_{n,t+h} \leftarrow \bar z^*$, and removes the new saturated commodities from the set of unsaturated commodities, i.e., $\unsaturated\setminus (n,h)$. Similarly, in each subsequent iteration $k$, Algorithm~\ref{alg.Generalized_MCFalgorithm} finds the \emph{lowest} admission rate $\bar z^*$ that saturates at least one \emph{unsaturated} commodity $(n,h)\in\unsaturated$, assigns $z_{n,t+h} \leftarrow \bar z^*$, and performs $\unsaturated\setminus (n,h)$.  The algorithm terminates when all commodities are saturated, i.e. $\unsaturated=\emptyset$. 

Consider one of the commodities $(n,h)$ that became saturated during iteration $k$. To increase its admission rate beyond saturation $z_{n,t+h}$, we would have to reduce the admission rate of at least one other commodity $(n',h)$ that became saturated either in iteration $k$ or in a previous iteration\footnote{Notice that if we could increase the admission rate of $(n,h)$ beyond saturation $z_{n,t+h}$ without reducing the admission rates of another saturated commodity $(n',h)$, then $(n,h)$ was not saturated.}. Notice that, by the design of Algorithm~\ref{alg.Generalized_MCFalgorithm}, the saturation admission rate of commodity $(n',h)$ is lower or equal to $z_{n,t+h}$. This means that, in each iteration $k$, the set of saturated admission rates $\{z_{n,t+h}\}_{(n,h)\notin\unsaturated}$ is max-min fair. It follows that, upon termination, Algorithm~\ref{alg.Generalized_MCFalgorithm} yields admission rates $\{z_{n,t+h}\}_{n=1}^{N-1}$ that are max-min fair.
\end{IEEEproof}

\section{Performance Evaluation} \label{sec.Evaluation}
In this section, we evaluate the performance of the PNR framework. In particular, in Sec.~\ref{sec.evalAP} we evaluate the prediction accuracy of the AP mechanism and compare it with two benchmark time series prediction methods. Then, in Sec.~\ref{sec.ToyNetwork}, we evaluate the performance of the MSNR algorithm and compare it with two reactive algorithms using a small network with $N=3$ nodes, synthetically generated attenuation levels $x_{t}^{(k,l)}$ and synthetically generated attenuation predictions $\hat{x}_{t+h}^{(k,l)}$ with different (adjustable) prediction accuracies. The goal is to draw insight from this small and controllable setting. Finally, in Sec.~\ref{sec.RealNetwork}, we evaluate the PNR framework (with both the AP mechanism and the MSNR algorithm) using the backhaul network with $N=13$ nodes illustrated in Fig.~\ref{CMLSweden} and the attenuation measurements from the dataset.

\subsection{Evaluation of the AP mechanism}\label{sec.evalAP}
The prediction accuracy of the AP mechanism is evaluated using the test sequences of attenuation measurements described in Sec.~\ref{sec.Model}. In this section, we show the results associated with \emph{Test Seq.~I} and \emph{Test Seq.~II}, both of which include a period of rain. We first assess the prediction error of a given link, then we analyze the prediction RMSE of the entire network and, finally, we assess the empirical probability of large prediction errors. 

\begin{figure}[t] 
\centering
\subfloat[]
{\includegraphics[width=0.3\columnwidth]{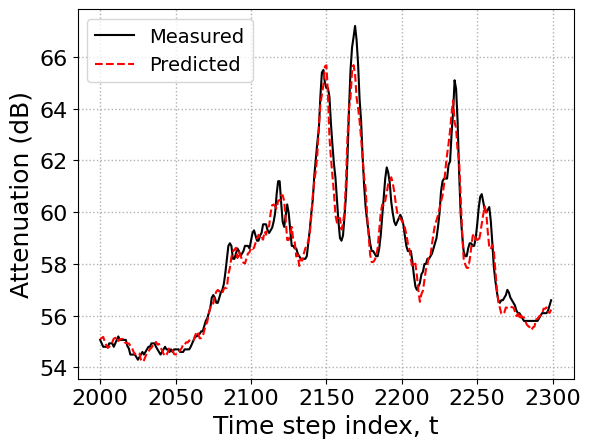}}
\hspace{0.5cm}
\subfloat[]
{\includegraphics[width=0.35 \columnwidth]{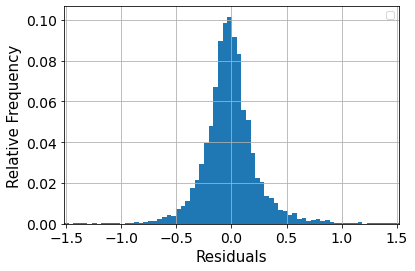}}
\label{hist}
\caption{(a) Comparison of the attenuation measurements from link $(9,13)$ with the corresponding  $3$-steps-ahead predictions. (b) Relative frequency distribution of the 3-steps-ahead prediction error.
\vspace{-0.5cm}
}\label{fig.predic_error}
\end{figure}

Let $e_{t,h}^{(k,l)} = x_{t+h}^{(k,l)} -\hat{x}_{t+h}^{(k,l)}$ be the $h$-\emph{steps-ahead prediction error} associated with link $(k,l)$ at time~$t$. In Fig.~\ref{fig.predic_error}(a), we compare the evolution of the attenuation measurements $x_{t+3}^{(9,13)}$ from link $(9,13)$ with the $3$-steps-ahead attenuation predictions $\hat{x}_{t+3}^{(9,13)}$ generated by the AP mechanism during an interval of $300$ time-steps from \emph{Test Seq.~I}. In Fig.~\ref{fig.predic_error}(b), we display the relative frequency distribution of the 3-steps-ahead prediction error $e_{t,3}^{(9,13)}$ from link $(9,13)$ associated with the entire \emph{Test Seq.~I}. The results in Fig.~\ref{fig.predic_error} suggest that: (i) the attenuation predictions accurately track the measurements and (ii) the distribution of the prediction error $e_{t,h}^{(k,l)}$ is similar to a normal distribution with zero mean.

\begin{figure}[t]
\centering
\subfloat[Test Seq.~I]{\includegraphics[width=0.35\columnwidth]{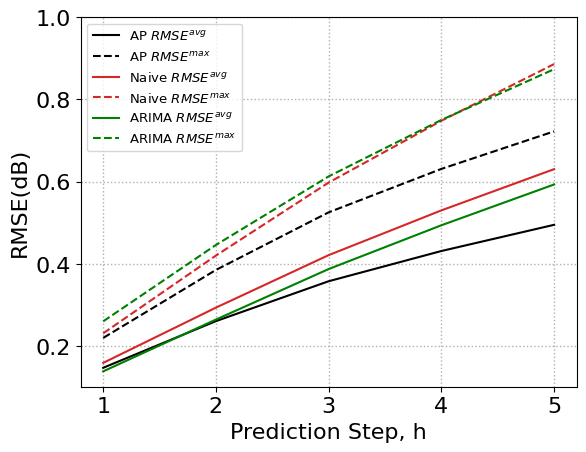}}
\hspace{0.5cm}
\subfloat[Test Seq.~II]{\includegraphics[width=0.35\columnwidth]{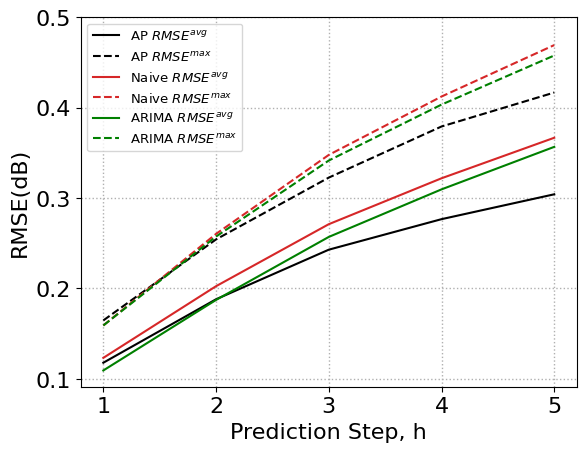}}
\caption{$RMSE_h^{\textrm{avg}}$ and $RMSE_h^{\textrm{max}}$ of the prediction error for different prediction horizons $h$ and for the AP mechanism, naive AP method, and ARIMA model.}\label{fig.RMSE_perc1}
\vspace{-0.5cm}
\end{figure}

Weather-induced attenuation varies over time and geographical location, and also depends on link's characteristics such as frequency, polarization, and length, meaning that prediction errors may differ considerably across different links. To capture the prediction error in the entire network, we employ 
\begin{align}
RMSE_{h}^{\textrm{avg}} &= \sqrt{\frac{1}{T-H}\sum_{t=1}^{T-H} \frac{1}{|E|}\sum_{{(k,l)} \in E} \left(e^{(k,l)}_{t,h}\right)^{2}} \\ RMSE_{h}^{\textrm{max}} &= \sqrt{\frac{1}{T-H}\sum_{t=1}^{T-H} \max_{{(k,l)}\in E}\left\{\left(e^{(k,l)}_{t,h}\right)^{2}\right\}} 
\end{align}
which calculate the RMSE associated with the $h$-steps-ahead prediction errors of all links over the entire time-horizon and the RMSE associated with the \emph{largest} $h$-steps-ahead prediction error among all the links in each time-step~$t$, respectively. 
In Fig.~\ref{fig.RMSE_perc1}, we display the $RMSE_h^{\textrm{avg}}$ and $RMSE_h^{\textrm{max}}$ (in dB) as a function of the prediction horizon $h\in\{1,\ldots,H\}$ for \emph{Test Seq.~I} and \emph{II} for three prediction mechanisms: (i) the AP mechanism; (ii) the naive AP method, also called random-walk method, which is a commonly used benchmark \cite{hyndman2006another} that employs the latest measurement as future predictions, namely $\hat{x}^{(k,l)}_{t+h}(naive) = x^{(k,l)}_t,\forall h$; and (iii) the ARIMA model, which is a well-known time series prediction model. 
For an example of the ARIMA model being employed to predict rain-induced attenuation in Ku-band  satellite links, we refer the reader to~\cite{patel2014implementation}. \emph{It is important to emphasize that both benchmark methods (i.e., naive and ARIMA) consider each link in isolation when predicting future attenuation levels and, thus, they do not capture the spatial correlation that is typical of weather-induced attenuation.}
The results in Fig.~\ref{fig.RMSE_perc1} suggest that the AP mechanism outperforms the benchmark methods in both \emph{Test Seqs.~I} and \emph{II} and that this performance improvement increases as the prediction horizon $h$ increases. In particular, when $h=1$, the performance improvement (in terms of $RMSE_{h}^{\textrm{avg}}$) of employing the AP mechanism as opposed to any of the benchmark methods is between $0\%$ and $12\%$, and when $h=5$, the performance improvement is between $12\%$ and $34\%$.

\begin{figure}[t]
\centering
\subfloat[Test Seq.~I]{\includegraphics[width=0.35\columnwidth]{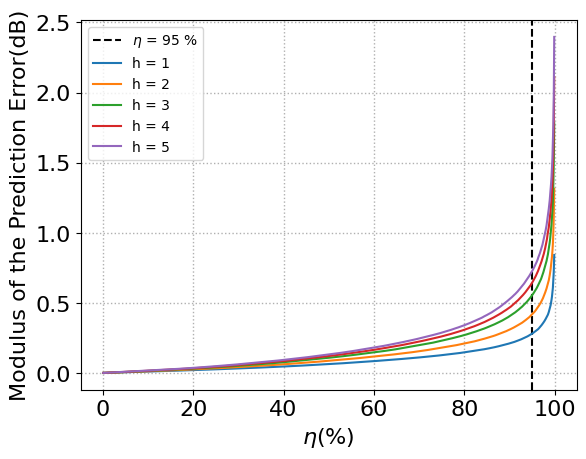}}
\hspace{0.5cm}
\subfloat[Test Seq.~II]{\includegraphics[width=0.35\columnwidth]{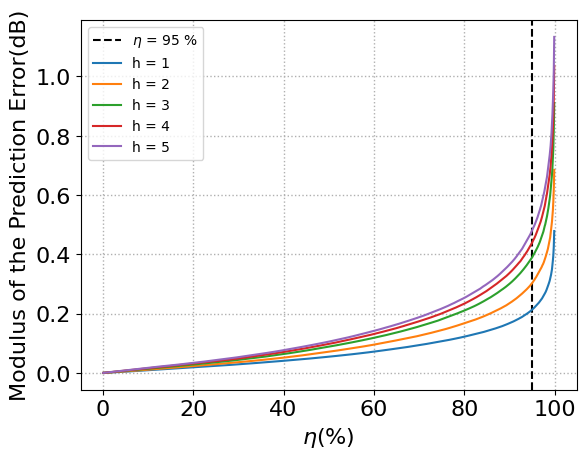}}
\caption{Percentile for the modulus of the prediction error.}\label{fig.RMSE_perc2}
\vspace{-0.5cm}
\end{figure}

To analyze the empirical probability of large prediction errors, we compute the percentile associated with the modulus of the $h$-steps-ahead predictions errors $|e^{(k,l)}_{t,h}|$. In particular, for a given test sequence with attenuation measurements $x_{t+h}^{(k,l)}$ and associated predictions $\hat{x}_{t+h}^{(k,l)}$ from the AP mechanism, the $\eta$\thinspace{th} percentile value represents the lowest $|e^{(k,l)}_{t,h}|$ that is larger than or equal to $\eta\%$ of all the values of $|e^{(k,l)}_{t,h}|$ in the considered dataset. 
For example, if the $95$\thinspace{th} percentile value for \emph{Test Seq.~I} and $h=3$ is $1$\thinspace{dB}, it means that $95\%$ of all the values of $|e^{(k,l)}_{t,h}|$ computed for the entire \emph{Test Seq.~I} are lower than or equal to $1$\thinspace{dB}. In Fig.~\ref{fig.RMSE_perc2}, we show the percentile values for different prediction horizons $h\in\{1,\ldots,5\}$ for \emph{Test Seq.~I} and \emph{II}. The results in Fig.~\ref{fig.RMSE_perc2} suggest that, as expected, the percentile values increase with the prediction horizon $h$. Moreover, the results show that $95\%$ of the one-step-ahead and $5$-steps-ahead prediction errors are lower than $0.5$\thinspace{dB} and $1.5$\thinspace{dB}, respectively.

In summary, the results in Figs.~\ref{fig.predic_error}, \ref{fig.RMSE_perc1}, and \ref{fig.RMSE_perc2} show that \emph{the AP mechanism predicts future link attenuation with high accuracy}. Next, we show that prediction accuracy has a significant impact on the performance of the MSNR algorithm.

\subsection{Evaluation of reconfiguration algorithms in a small and controllable network}\label{sec.ToyNetwork}
We compare the performance of the \textsc{MSNR} algorithm with two reactive network reconfiguration algorithms, namely \textsc{Never re-route} and \textsc{Always re-route}, in terms of their network utilization, which is captured by the evolution of the node-average admission rate $\sum_{n=1}^{N-1}z_{n,t}/(N-1)$ over time~$t$. 
The considered network reconfiguration algorithms are:
\begin{itemize}
    \item[i)] \emph{\textsc{MSNR} algorithm}: leverages future predicted capacities to decide when to re-route. In particular, in each time-step~$t$, it compares the performance of different re-routing plans and selects the plan $\mathbf{\route_t^{*}}$ with highest cumulative sum of admission rates, as 
    described in Sec.~\ref{sec.Multple}. 
    \item[ii)] \emph{\textsc{Never re-route} algorithm}: attempts to maximize the admission rates $z_{n,t}$ by never provisioning scratch capacity and, thus, fully utilizing links whenever possible. This reactive algorithm operates based on the MSNR algorithm. However, instead of selecting $\mathbf{\route_t^{*}}$, it selects, in \emph{every} time-step~$t$, the re-routing plan $\mathbf{\route_t}=(0,\ldots,0)$. Under this algorithm, the SDN controller is rarely\footnote{Notice that if the predicted capacities $\hat c_{t+1}^{(k,l)}$ are inaccurate, in particular if $\hat c_{t+1}^{(k,l)} < c_{t+1}^{(k,l)}$, it may happen that $\scratch_{t+1}\geq\scratch_{min}$ and the \textsc{Never re-route} algorithm is allowed to re-route at time~$t+1$.} allowed to re-route, but it is continually optimizing the admission rates. 
    \item[iii)] \emph{\textsc{Always re-route} algorithm}: attempts to provision scratch capacity $\scratch_{t}\geq\scratch_{min}=0.05$ at every time-step~$t$, enabling the SDN controller to optimize routing decisions often. This reactive algorithm operates based on the MSNR algorithm. However, instead of selecting $\mathbf{\route_t^{*}}$, it selects, in \emph{every} time-step~$t$ with $\scratch_{t}\geq\scratch_{min}$, the re-routing plan $\mathbf{\route_t}=(1,1,\ldots,1)$, and in \emph{every} time-step~$t$ with $\scratch_{t}<\scratch_{min}$, the re-routing plan $\mathbf{\route_t}=(0,1,\ldots,1)$. 
\end{itemize}
Notice that all three network reconfiguration algorithms select max-min fair admission rates $z_{n,t}$ \emph{at every time-step} $t$. The main difference between them is that only the \textsc{MSNR} algorithm employs the predictions of the links' future condition to decide when to re-route. Both the \textsc{Never re-route} and \textsc{Always re-route} algorithms simply react to the time-varying conditions of the network. The comparison with the predictive SDN-based routing framework developed in \cite{Yaghoubi2018} is not possible due to the incompatible assumptions. Recall that the framework in \cite{Yaghoubi2018} can \emph{only} be employed during periods of rain, it allows flows to temporarily exceed the link capacity, and it does not take fairness into account, thus, making the comparison unfit. Next, we evaluate the three reconfiguration algorithms in a small and controllable setting using synthetic data. In Sec.~\ref{sec.RealNetwork}, we evaluate the same three algorithms using measurements collected from the backhaul network.

The results in this section are associated with the network in Fig.~\ref{fig.ToyNetwork} with $N=3$ nodes and three links $\{(1,2),(2,3),(1,3)\}$. The \emph{normalized}\footnote{Both demands and capacities are normalized with respect to the maximum achievable bitrate of 225\thinspace{Mbps} from Table~\ref{tab.adaptiveModulation}.} demands associated with nodes $1$ and $2$ remain fixed at $d_1=1$ and $d_2=0.5$, respectively, during the time-horizon of $1,000$ time-steps. The (actual) attenuation levels $x_t^{(k,l)}$ and predicted attenuation levels $\hat x_{t+h}^{(k,l)}$ are synthetically generated according to the following stochastic processes
\begin{align}
    x_{t}^{(k,l)} &= \min\{ \max\{ x_{t-1}^{(k,l)} + \error_{t}^{(k,l)} ; -100 \} ; -50 \}\; ; \label{eq.curr_capc}\\
    \hat x_{t+h}^{(k,l)} &= \min\{ \max\{ x_{t+h}^{(k,l)} + \tilde \error_{t,h}^{(k,l)} ; -100 \} ; -50 \}\; , \label{eq.future_capc}
\end{align}
for all links $(k,l)\in E$, for all time-steps $t\in\{1,\ldots,1,000\}$, for all values of $h\in\{1,\ldots,H\}$, and with $x_{0}^{(k,l)}$ sampled from a  uniform distribution in the interval $(-100,-50)$. Notice that \eqref{eq.curr_capc} establishes the variation of the attenuation $x_{t}^{(k,l)}$ over time, while \eqref{eq.future_capc} establishes the noise in the prediction $\hat x_{t+h}^{(k,l)}$ of the future attenuation $x_{t+h}^{(k,l)}$. 
The sequence of Gaussian random variables $\error_{t}^{(k,l)}$ is i.i.d.\ over time~$t$, independent across links, and sampled according to $\mathcal{N}(0,6.25)$.  
Similarly, the sequence of random variables $\tilde \error_{t,h}^{(k,l)}$ are Gaussian $\mathcal{N}(0,\tilde\sigma^2)$ with positive variance $\tilde\sigma^2$, i.i.d.\ over time, and independent across different links. 
Notice from \eqref{eq.future_capc} that, a high variance $\tilde\sigma^2$ represents an AP mechanism with poor accuracy, i.e. large prediction error. The choice of Gaussian distribution for $\tilde \error_{t,h}^{(k,l)}$ was inspired by the relative frequency distribution of the prediction error shown in Fig.~\ref{fig.predic_error}(b). 

\begin{figure}[t]
  \centering
  \subfloat[Normalized link capacity $c_{t}^{(k,l)}$]{\includegraphics[width=0.32\columnwidth]{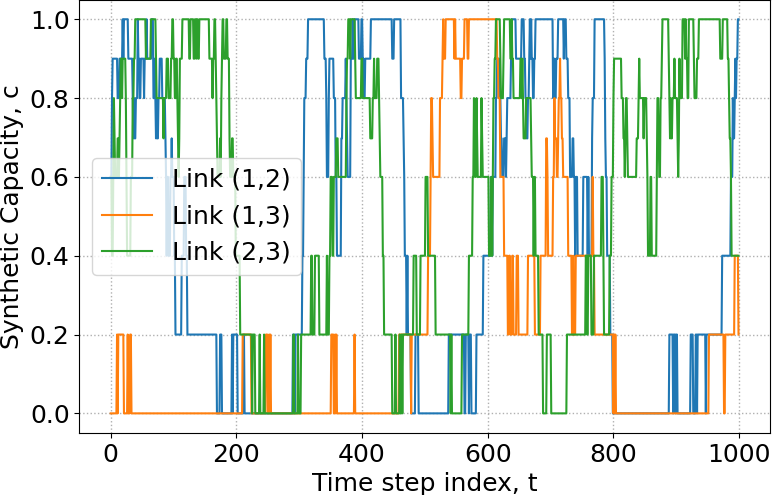}}
  \hfill
  \subfloat[Node-aver.\ rate $(z_{1,t}+z_{2,t})/2$]{\includegraphics[width=0.32\columnwidth]{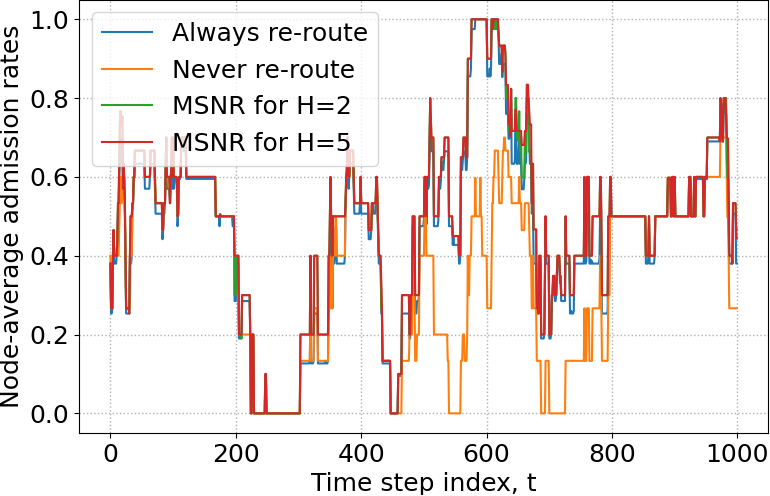}}
  \hfill
  \subfloat[Time-aver.\ $\sum_{t=1}^T(z_{1,t}+z_{2,t})/2T$]{\includegraphics[width=0.32\columnwidth]{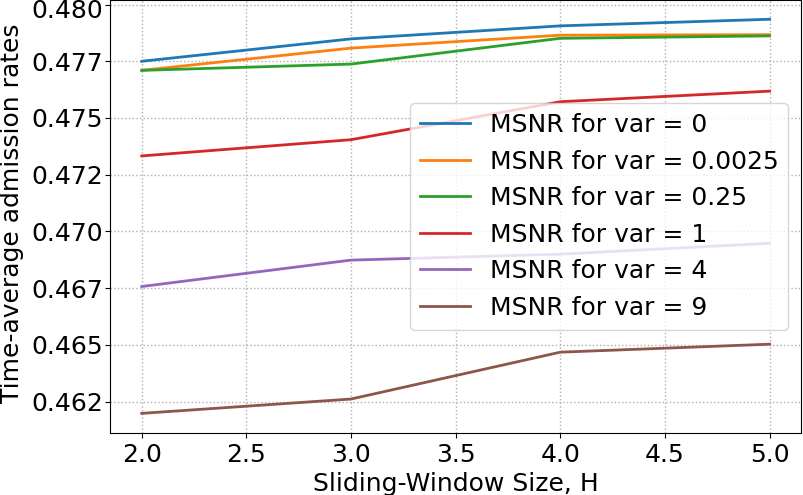}}
  \caption{Performance of the MSNR algorithm for the network in Fig.~\ref{fig.ToyNetwork} with $N=3$ nodes. 
  (a) Evolution of the normalized link capacity $c_{t}^{(k,l)}$ over time. 
  (b) Node-average admission rate $(z_{1,t}+z_{2,t})/2$ for different reconfiguration algorithms with ideal attenuation prediction ($\tilde\sigma^2=0$).
  (c) Time-average admission rate $\sum_{t=1}^T(z_{1,t}+z_{2,t})/2T$ for MSNR with different prediction window sizes $H\in\{2,3,4,5\}$ and attenuation prediction accuracies $\tilde\sigma^2\in\{0,0.0025,\ldots,9\}$.} \label{fig.toy_experiment}
  \vspace{-0.5cm}
\end{figure}

To determine the (actual) capacities $c_{t}^{(k,l)}$ and the predicted capacities $\hat c_{t+h}^{(k,l)}$ associated with the synthetic values of $x_{t}^{(k,l)}$ and $\hat x_{t+h}^{(k,l)}$, respectively, we adopt a constant transmission signal level of $P_{Tx,t}^{(k,l)}=0$\thinspace{dBm} and use the AM mechanism described in Sec.~\ref{sec.AMM}. In Fig.~\ref{fig.toy_experiment}(a), we display the evolution of the \emph{normalized} values of $c_{t}^{(k,l)}$ employed to obtain the results in this section. Notice that this is a network with highly dynamic link capacities $c_{t}^{(k,l)}$. 

In Fig.~\ref{fig.toy_experiment}(b), we compare the evolution of the node-average admission rate $(z_{1,t}+z_{2,t})/2$ over time~$t$ for different reconfiguration algorithms operating with ideal attenuation predictions, i.e., with $\hat x_{t+h}^{(k,l)}=x_{t+h}^{(k,l)}$ and, as a result, $\hat c_{t+h}^{(k,l)}=c_{t+h}^{(k,l)}$. In Fig.~\ref{fig.toy_experiment}(c) and in Table~\ref{tab.toy_network}, we show the time-average admission rates $\sum_{t=1}^T(z_{1,t}+z_{2,t})/2T$ for different reconfiguration algorithms operating with attenuation predictions with different accuracies $\tilde\sigma^2\in\{0,$ $0.0025,$ $0.25,$ $1,$ $4,$ $9,$ $25\}$ and different prediction window sizes $H\in\{2,3,4,5\}$. 


The results in Fig.~\ref{fig.toy_experiment}(b) show that, as expected, \textsc{Never re-route} has the worse performance, while MSNR with prediction window size $H=5$ has the best performance in terms of network utilization. The poor performance of \textsc{Never re-route}, especially between time-steps $500$ and $800$, results from the SDN controller not being allowed to re-route. The lower performance of \textsc{Always re-route} when compared to MSNR is due to the frequent provisioning of scratch capacity $\scratch_{min}=0.05$. By leveraging the prediction of links' future conditions, MSNR can assess the potential future benefits of re-routing\footnote{Recall from the discussion in Sec.~\ref{sec.Cost} that planning to re-route at the next time-step~$t+1$, can only hurt the network performance at the current time~$t$ due to the provision of the scratch capacity.}, which allows it to choose when is the best time to re-route. Throughout the $1,000$ time-steps, the SDN controller re-routes $31$, $30$, $28$, and $29$ times when employing MSNR with prediction window sizes $H\in\{2,3,4,5\}$, respectively.

The results in Fig~\ref{fig.toy_experiment}(c) and Table~\ref{tab.toy_network} suggest that: (i) the performance of MSNR improves as the prediction accuracy improves and as the window size $H$ increases 
and (ii) the performance gain of improving the prediction accuracy is more significant than the performance gain of increasing the prediction window size $H$, which \emph{highlights the importance of developing an accurate AP mechanism}.

\begin{figure}[t]
  \centering
  \subfloat[Normalized link capacity $c_{t}^{(k,l)}$]{\includegraphics[width=0.32\columnwidth]{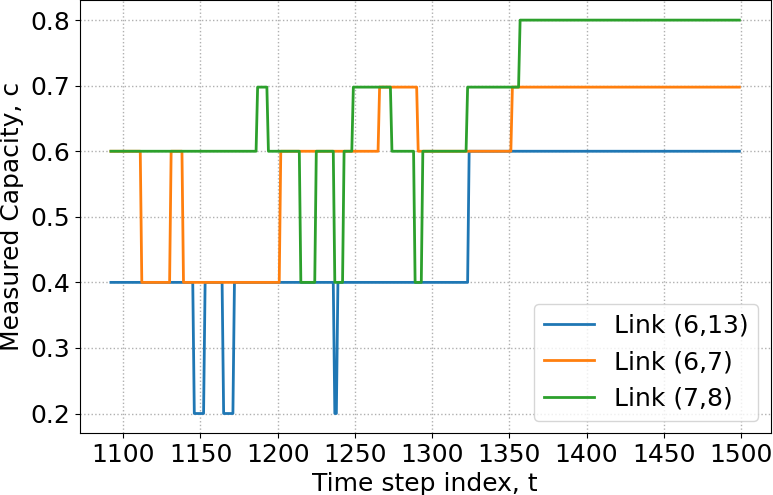}}
  \hfill
  \subfloat[Ideal attenuation prediction]{\includegraphics[width=0.32\columnwidth]{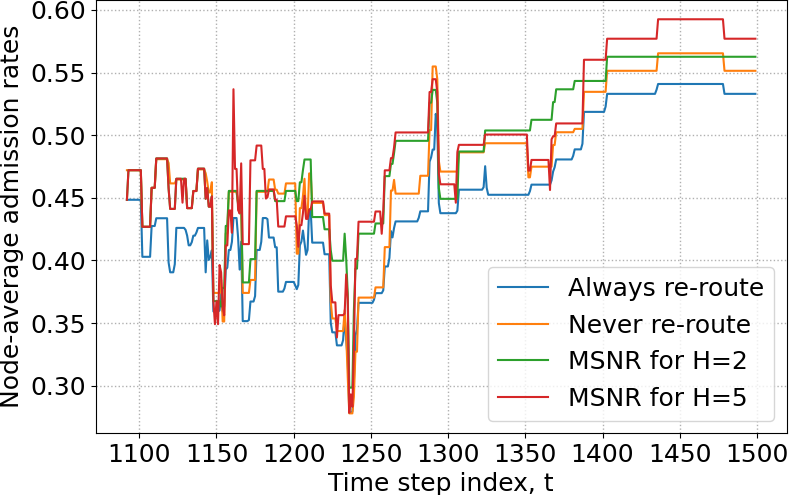}}
  \hfill
  \subfloat[Prediction from AP mechanism]{\includegraphics[width=0.32\columnwidth]{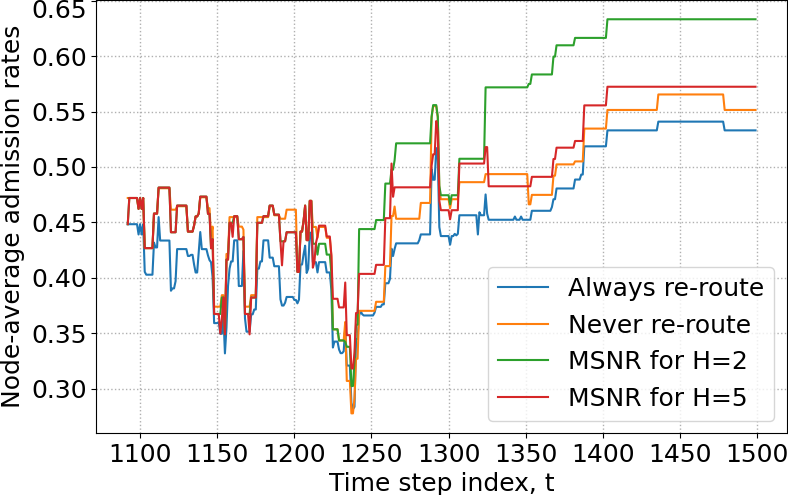}}
  \vspace{-0.2cm}
  \caption{
  Performance of the PNR framework using data collected from the backhaul network in Fig.~\ref{CMLSweden}. (a) Evolution of the normalized measured link capacity $c_{t}^{(k,l)}$ over time. (b)-(c): Evolution of the node-average admission rate $\sum_{n=1}^{N-1}z_{n,t}/(N-1)$ over time for different network reconfiguration algorithms.} \label{fig.real_experiment}
  \vspace{-0.5cm}
\end{figure}

\begin{table}[t]
\caption{Time-average admission rates $\sum_{t=1}^T(z_{1,t}+z_{2,t})/2T$ for different network reconfiguration algorithms and for attenuation predictions with different accuracies.}\label{tab.toy_network}\vspace{-0.3cm}
\begin{center}
\begin{tabular}{cccc} \toprule
Prediction accuracy & Ideal & $\tilde\sigma^2=1$ & $\tilde\sigma^2=25$ \\
\midrule
\textsc{Never re-route} & 0.362 & 0.359 & 0.343 \\
\textsc{Always re-route} & 0.453 & 0.449 & 0.427 \\
MSNR for $H=2$ & 0.477 & 0.473 & 0.447 \\
MSNR for $H=3$ & 0.478 & 0.474 & 0.449 \\
MSNR for $H=4$ & 0.479  & 0.476 & 0.449 \\
MSNR for $H=5$ & 0.479 & 0.476 & 0.450 \\
\bottomrule
\end{tabular}
\vspace{-0.5cm}
\end{center}
\end{table}


\begin{table}[t]
\caption{Performance gain of MSNR with $H\in\{2,5\}$ when compared to a reactive algorithm: \textsc{Never re-route} or \textsc{Always re-route}. The performance gain metrics in columns 3, 4, and 5 are defined in \eqref{eq.time_average}.}\label{tab.real_network}\vspace{-0.3cm}
\begin{center}
\begin{tabular}{ccccc} \toprule
MSNR & Reactive &  Time-average & Node-average & Instantaneous \\
\midrule
$H=5$ & \textsc{Always} & $7.74\%$ & $18.00\%$ & $170.19\%$ \\
$H=2$ & \textsc{Always} & $15.49\%$ & $26.84\%$ & $263.58\%$ \\
$H=5$ & \textsc{Never} & $1.67\%$ & $10.37\%$ & $68.37\%$ \\
$H=2$ & \textsc{Never} & $8.98\%$ & $22.04\%$ & $208.01\%$ \\
\bottomrule
\end{tabular}
\vspace{-0.5cm}
\end{center}
\end{table}

\subsection{Evaluation of the PNR Framework with data from a real-world network}\label{sec.RealNetwork}
We now evaluate the performance of the PNR framework using the data collected from the backhaul network in Fig.~\ref{CMLSweden} with $N=13$ base-stations ($12$ commodities and one destination) and $17$ links. The normalized demands assigned to the commodities are chosen according to a uniform distribution in the interval $(0,2)$. 
In particular, the twelve demand values\footnote{Notice that similar results can be obtained for different vectors of demands.} are $\boldsymbol{d}=[1.111,$ $0.557,$ $1.124,$ $1.266,$ $0.174,$ $1.485,$ $0.947,$ $0.067,$ $0.140,$ $0.596,$ $1.413,$ $0.999]$.
The values of the (actual) capacities $c_{t}^{(k,l)}$ and future predicted capacities $\hat{c}_{t+h}^{(k,l)}$ are determined by the link attenuation measurements in the dataset, 
by the AM mechanism described in Sec.~\ref{sec.AMM}, and by the AP mechanism. To train, tune, and test the AP mechanism, we use a train-validation-test split of 80-10-10. To assess the performance of the PNR framework in a \emph{challenging scenario}, we choose a sequence of more than $400$ measurements (from \emph{Test Seq.~I} described in Sec.~\ref{sec.Model}) that includes a period with high attenuation variability due to a rain event. Moreover, we consider transmission signal levels $P_{Tx,t}^{(k,l)}$ that are $10$\thinspace{dBm} lower than the dataset measurements. In Fig.~\ref{fig.real_experiment}(a), we display the evolution of the normalized capacities $c_{t}^{(k,l)}$ from three selected links. Notice that the variation is significant. In Fig.~\ref{fig.real_experiment}(b), we show the evolution of the node-average admission rates $\sum_{n=1}^{N-1}z_{n,t}/(N-1)$ for different reconfiguration algorithms employing ideal attenuation prediction, i.e., $\hat c_{t+h}^{(k,l)}=c_{t+h}^{(k,l)}$. In Fig.~\ref{fig.real_experiment}(c), we display the node-average admission rates for algorithms employing the AP mechanism to predict $\hat c_{t+h}^{(k,l)}$ over time.  
The results in Figs.~\ref{fig.real_experiment}(b) and \ref{fig.real_experiment}(c) show that MSNR outperforms both \textsc{Never re-route} and \textsc{Always re-route}. 

In Table~\ref{tab.real_network}, we display the performance gain of MSNR with $H\in\{2,5\}$ employing the AP mechanism when compared to reactive algorithms: \textsc{Never re-route} or \textsc{Always re-route}. Let $z^{(M)}_{n,t}$ and $z^{(R)}_{n,t}$ be the admission rates associated with MSNR and the reactive algorithm, respectively. 
The third, fourth, and fifth columns of Table~\ref{tab.real_network} are associated with
\begin{equation}\label{eq.time_average}
\frac{\sum_{t=1}^T\sum_{n=1}^{N-1}(z^{(M)}_{n,t}-z^{(R)}_{n,t})}{\sum_{t=1}^T\sum_{n=1}^{N-1}z^{(R)}_{n,t}} \; \mbox{ , } \;  \max_{t}{\left\{\frac{\sum_{n=1}^{N-1}(z^{(M)}_{n,t}-z^{(R)}_{n,t})}{\sum_{n=1}^{N-1}z^{(R)}_{n,t}}\right\}} \; \mbox{ , } \;  \max_{n,t}{\left\{\frac{z^{(M)}_{n,t}-z^{(R)}_{n,t}}{z^{(R)}_{n,t}}\right\}} \; ;
\end{equation}
which represent the \emph{time-average performance gain}, the \emph{maximum node-average performance gain}, and the \emph{maximum instantaneous performance gain}, respectively. 
The results in Table~\ref{tab.real_network} show that the MSNR algorithm can improve the time-average admission rate $\sum_{t=1}^T(z_{1,t}+z_{2,t})/2T$ by more than $7\%$ when compared to either \textsc{Always re-route} or \textsc{Never re-route} and, more importantly, they also show that the gain in terms of the instantaneous per commodity admission rate $z_{n,t}$ can exceed $200\%$. These significant instantaneous gains occur when severe rain-induced attenuation occurs, showing that the PNR framework is able to prepare the network ahead of time and alleviate the impact of these severe disturbances on the network performance, which can be paramount to time-sensitive applications.

An important observation from the results in Secs.~\ref{sec.ToyNetwork} and \ref{sec.RealNetwork} is that, when the AP mechansism has high accuracy, the performance gap between MSNR with $H=2$ and reactive algorithms is significantly larger than the performance gain obtained from increasing the prediction window size $H$. Adding to this observation the fact that the computational complexity of MSNR grows with $H$, as discussed in Sec.~\ref{sec.Multple}, makes the PNR framework with $H=2$ an attractive choice both in terms of performance and complexity. 

\section{Conclusion}\label{sec.Conclusion}
We developed the PNR framework, that includes: (i) the AP mechanism that uses historical data to predict the sequence of future attenuation levels, \emph{without incorporating any specific weather-related models}; and (ii) the MSNR algorithm that dynamically optimize routing $f_{n,t}^{(k,l)}$ and admission control $z_{n,t}$ decisions over time aiming to maximize the cumulative sum of admission rates $\sum_{t=1}^T\sum_{n=1}^{N-1}z_{n,t}$, while ensuring that, in every time-step $t$, the selected feasible set $\{f_{n,t}^{(k,l)},z_{n,t}\}$ is max-min fair in every time-step $t$ and can be implemented without inducing transient congestion. We use a real-world dataset to thoroughly evaluate the PNR framework and to show that it allows the SDN controllers to prepare the x-haul for imminent (and possibly severe) weather-induced disturbances. 
%
%
%
There are several open problems that will be considered in our future work, including consideration of time-varying traffic demands $d_n$, consideration of downlink/uplink traffic, application to 5G slice admission and provisioning, and experimental evaluation in city-scale testbeds.

\section{Acknowledgement}
This work was supported in part by the NSF-BSF grant CNS-1910757 and the NSF grant OAC-2029295.


\bibliographystyle{IEEEtran}
\bibliography{bibtex/bib/AllRef}

\end{document}